\documentclass[dvipsnames]{aa}  

\usepackage{natbib}
\usepackage{graphicx}
\usepackage{txfonts}
\usepackage{hyperref}
\usepackage{listings}
\usepackage{amsmath}
\usepackage{tikz}
\usepackage{array}
\usepackage{soul}
\usepackage{colortbl}
\usepackage{xcolor}
\usepackage{multirow}
\usepackage{xparse}
\usepackage{bm}
\usepackage{url}
\newcommand{\changeAdd}[1]{#1}

\usetikzlibrary{calc,matrix,patterns,shadings,backgrounds}
\pgfdeclarelayer{myback}
\pgfsetlayers{myback,background,main}

\tikzset{myfillcolor/.style = {fill=#1}}%

\NewDocumentCommand{\highlight}{O{blue!40} m m}{%
\draw[myfillcolor=#1] (#2.north west)rectangle (#3.south east);
}

\NewDocumentCommand{\vshade}{O{blue!40} O{white} m m}{%
\draw[bottom color =#1,top color=#2] (#3.north west)rectangle (#4.south east);
}

\NewDocumentCommand{\oshade}{O{blue!40} O{white} m m}{%
\draw[right color =#1,left color=#2] (#3.north west)rectangle (#4.south east);
}

\NewDocumentCommand{\inshade}{O{blue!40} O{white} m m}{%
\draw[inner color =#1,outer color=#2] (#3.north west)rectangle (#4.south east);
}

\NewDocumentCommand{\fillpattern}{O{north west lines} O{blue!50} m m}{%
\draw[pattern=#1, pattern color=#2] (#3.north west)rectangle (#4.south east);
}

\definecolor{bittersweet}{rgb}{0.75, 0.58, 0.89}
\newcommand{\colsquare}[1]{\cellcolor{#1}\color{#1}.}

\def\checkmark{\tikz\fill[scale=0.4](0,.35) -- (.25,0) -- (1,.7) -- (.25,.15) -- cycle;} 

\def\zs{\mbox{{$z_{\rm spec}$}}}
\def\zp{\mbox{{$z_{\rm phot}$}}}

\begin{document} 

  \title{Return of the features}
  \subtitle{Efficient feature selection and interpretation for photometric redshifts}
  \author{A.~D'Isanto\inst{\ref{inst1}} \and S.~Cavuoti\inst{\ref{inst2},\ref{inst3},\ref{inst3b}}\and F.~Gieseke\inst{\ref{inst4}}\and K.L.~Polsterer\inst{\ref{inst1}}}

  \institute{Astroinformatics group, Heidelberg Institute for Theoretical Studies,
  Schloss-Wolfsbrunnenweg 35, 69118 Heidelberg, Germany \label{inst1}\\
    \email{antonio.disanto@h-its.org},
    \email{kai.polsterer@h-its.org}
    \and 
    Department of Physics ``E. Pancini'', University Federico II, via Cinthia 6, I-80126, Napoli, Italy
     \label{inst2}
    \and 
    INAF - Astronomical Observatory of Capodimonte, via Moiariello 16, 80131 Napoli, Italy \label{inst3} \and
    INFN - Section of Naples, via Cinthia 9, 80126 Napoli, Italy \label{inst3b}
    \email{stefano.cavuoti@gmail.com}
    \and Machine Learning Group Image Section, Department of Computer Science, University of Copenhagen, Sigurdsgade 41, 2200 København N, Denmark \label{inst4}
    }

  \date{Received }

\abstract{
    The explosion of data in recent years has generated an increasing need for new analysis techniques in order to extract knowledge from massive data-sets.
    Machine learning has proved particularly useful to perform this task.
    Fully automatized methods (e.g. deep neural networks) have recently gathered great popularity, even though those methods often lack physical interpretability.
    In contrast, feature based approaches can provide both well-performing models and understandable causalities with respect to the correlations found between features and physical processes.
}{
    Efficient feature selection is an essential tool to boost the performance of machine learning models.
    In this work, we propose a forward selection method in order to compute, evaluate, and characterize better performing features for regression and classification problems.
    Given the importance of photometric redshift estimation, we adopt it as our case study.
}{
    We synthetically created $4,520$ features by combining magnitudes, errors, radii, and ellipticities of quasars, taken from the Sloan Digital Sky Survey (SDSS).
    We apply a forward selection process, a recursive method in which a huge number of feature sets is tested through a k-Nearest-Neighbours algorithm, leading to a tree of feature sets.
    The branches of the feature tree are then used to perform experiments with the random forest, in order to validate the best set with an alternative model.
}{
    We demonstrate that the sets of features determined with our approach improve the performances of the regression models significantly when compared to the performance of the classic features from the literature.
    The found features are unexpected and surprising, being very different from the classic features.
    Therefore, a method to interpret some of the found features in a physical context is presented.
}{
    The feature selection methodology described here is very general and can be used to improve the performance of machine learning models for any regression or classification task.
}
\maketitle

\section{Introduction}
In recent years, astronomy has experienced a true explosion in the amount and complexity of the available data.
The new generation of digital surveys is opening a new era for astronomical research, characterized by the necessity to analyse data-sets that fall into the Tera-scale and Peta-scale regime.
This is leading to the need for a completely different approach with respect to the process of knowledge discovery.
In fact, the main challenge will no longer be obtaining data in order to prove or disprove a certain hypothesis, but rather to mine the data in order to find interesting trends and unknown patterns.
The process of discovery will not be driven by new kinds of instrumentation to explore yet unobserved regimes, but by efficient combination and analysis of already existing measurements.
Such an approach requires the development of new techniques and tools in order to deal with this explosion of data, which are far beyond any possibility of manual inspection by humans.
This necessity will become urgent in the next years, when surveys like the Large Synoptic Survey Telescope \citep[LSST][]{ivezic2008lsst}, the Square Kilometer Array \citep[SKA][]{2008IAUS..248..164T}, and many others, will become available.
Therefore, machine learning techniques are becoming a necessity in order to automatize the process of knowledge extraction from big data-sets.
In the last decade, machine learning has proved to be particularly useful to solve astrophysical complex non-linear problems, both for regression \citep[see for instance][]{Hildebrandt2010, 2014ApJS..210....9B, Cavuoti201545, Hoyle201634, Beck20174323} and classification tasks \cite[see][]{ 2008AN....329..288M, 2012MNRAS.427.2917R, Cavuoti2013968, disanto2016, Smirnov20172024, Benavente2017}.
These techniques find nowadays many applications in almost all the fields of science and beyond \citep{noauthororeditorfourth}.
In the literature, two main machine learning branches can be found that deal with the selection of the most relevant information contained in the data.
The first traditional way consists in the extraction and selection of manually crafted features, which are theoretically more suitable to optimize the performance.
In \citet{2013arXiv1310.1976D} feature selection strategies are compared in an astrophysical context.

The second option is using automatic feature selection models and became more popular in more recent years.
For example, \citet{2015arXiv150702313A} delegate this task to the machine by analysing the automatically extracted feature representations of convolutional neural networks.
In convolutional neural networks, during the training phase the model itself determines and optimizes the extraction of available information in order to obtain the best performance.
The challenge of feature selection is fundamental for machine learning applications, due to the necessity of balancing between overfitting and the curse of dimensionality \citep{Bishop:2006:PRM:1162264}, which arises when dealing with very high-dimensional spaces.
Therefore a clever process of feature selection is needed to overcome this issue.
In this setting, a different strategy was chosen for this work, in which a forward selection algorithm \citep{Guyon:2003:IVF:944919.944968} is adopted to identify the best performing features out of thousands of \changeAdd{them}.
We decided to apply this procedure in a very important field: photometric redshift estimation.
Due to the enormous importance that this measurement has in cosmology, great efforts have been lavished by the astronomical community on building efficient methods for the determination of affordable and precise photometric redshifts \citep{2001AJ....122.1151R,2008A&A...480..703H,2008ApJ...683...12B,Hildebrandt2010}.
Photometric redshifts are of extreme importance with respect to upcoming missions, for example the forthcoming Euclid mission \citep{2011arXiv1110.3193L}, which will be based on the availability of photometric redshift measures, and the Kilo Degree Survey \citep[KiDS][]{dejong2017}, which aims to map the large-scale matter distribution in the Universe, using weak lensing shear and photometric redshift measurements \citep{Hildebrandt20161,Tortora20162845,Harnois-Deraps20171619,Joudaki20171259,Kohlinger20174412}.
Furthermore, photometric redshifts estimation is crucial for several other projects, the most important being the Evolutionary Map of the Universe (EMU) \citep{2011PASA...28..215N}, the Low Frequency Array (LOFAR) \citep{2013A&A...556A...2V}, Dark Energy Survey \citep{2016PhRvD..94d2005B}, the Panoramic Survey Telescope \& Rapid Response System (PANSTARRS) \citep{2016arXiv161205560C}, and the VST Optical Imaging of the CDFS and ES1 Fields (VST-VOICE) \citep{2016heas.confE..26V}.
In light of this, we propose to invert the task of photometric redshift estimation.
That is to say, having stated the possibility to determine the redshift of a galaxy based on its photometry, we want to build a method that allows us to investigate the parameter space and to extract the features to be used to achieve the best performance.
As thoroughly analysed in \citet{2018A&A...609A.111D}, the implementation of deep learning techniques is providing an alternative to feature based methods, allowing the estimation of photometric redshifts directly from images.
The main concerns when adopting deep learning models are related to the amount of data needed to efficiently perform the training of the networks, the cost in terms of resources and computation time, and the lack of interpretability related to the features automatically extracted.
In fact, deep learning models can easily become like magic boxes and it is really hard to assign any kind of physical meaning to the features estimated by the model itself.
Therefore, a catalogue-based approach still has great importance, due to the gains in time, resources, and interpretability.
In particular, this is true if a set of significant features is provided, in order to concentrate the important information with respect to the problem in a reduced number of parameters.
Both methods, based on automatically extracted features or on selected features, constitute the starting point to build an efficient and performing model for redshift estimation, respectively.
The topic of feature selection is a well-treated subject in the literature \citep[see for example][]{2012MNRAS.427.2917R, Tangaro2015, 2015MNRAS.449.1275H, disanto2016}. 
The forward selection approach we used \citep{DBLP:conf_esann_GiesekePOI14} is meant to select between thousands of features generated by combining plain photometric features as they are given in the original catalogue.
No matter what selection strategy is applied, the final results have to be compared to those obtained with the traditional features from the literature \citep{2007ApJ...663..752D, 2009ApJS..180...67R, 2011MNRAS.418.2165L} and with automatically extracted features.
The aim is to find the subsets that give a better performance for the proposed experiments, mining into this new, huge feature space and to build a method useful to find the best features for any kind of problem.
Moreover, we propose to analyse the obtained features, in order to give them a physical explanation and a connection with the processes occurring in the specific category of sources.
Such an approach also demands a huge effort in terms of computational time and resources.
Therefore, we need an extreme parallelization to deal with this task.
This has been done through the intensive use of graphics processing units (GPU), a technology that is opening new doors for Astroinformatics \citep{Cavuoti201329,3578f9b7d115405c91ae0bcd686baaec,2018A&A...609A.111D}, allowing the adoption of deep learning and/or massive feature selection strategies.
In particular, in this work, the feature combinations are computed following \citet{DBLP:conf_esann_GiesekePOI14} and \citet{2014ASPC..485..425P}, using a GPU cluster equipped with four Nvidia Pascal P40 graphic cards.\footnote{\hfill \url{https://images.nvidia.com /content/pdf/tesla/184427- Tesla-P40-Datasheet-NV-Final-Letter-Web.pdf}.}
Likewise for \citet{2013AJ....146...22Z}, the k-Nearest-Neighbours \citep[kNN][]{fix_51_discriminatory} model is used, running recursive experiments in order to estimate the best features through the forward selection process.
This choice has been done because the kNN model scales very well with the use of GPU, with respect to performance and quality of the prediction, as shown in \citet{heinermann2013}.
In this way, for each run of the experiment, the most contributing features are identified and added to previous subsets.
Thereby, a tree of feature groups is created that afterwards can be compared with the traditional ones.
The validation experiments are performed using a random forest (RF) model \citep[application in astronomy]{2010ApJ...712..511C}.
We will show that this approach can strongly improve performance for the task of redshift estimation.
The improvement is due to the identification of specific feature subsets containing more information and capable of better characterizing the physics of the sources.
In the present work, we perform the experiments on quasar data samples extracted from the Sloan Digital Sky Survey Data Release 7 \citep[SDSS DR7][]{2009ApJS..182..543A} and Data Release 9 \citep[SDSS DR9][]{2012ApJS..203...21A}.
The proposed approach is very general and could be also used to solve many other tasks in astronomy, including both regression and classification problems.

\noindent
{\bf Outline:}
In Sec.~\ref{methods} the methodology and models used to perform the experiments are described together with the statistical estimators used to evaluate the performance.
The strategy adopted for the feature selection is also explained. 
Section~\ref{data} is dedicated to the data used and the feature extraction process.
In Sec.~\ref{results} the experiments performed and the results obtained are described.
Finally, in Sec.~\ref{discussion} the results are discussed in detail and in Sec.~\ref{conclusions} some conclusions are drawn.

\section{Methods}\label{methods}
The main purpose of this work is to build an efficient method capable of generating, handling and selecting the best features for photometric redshift estimation, even though the proposed method is also able to deal with any other task of regression or even classification.
We calculate thousands of feature combinations of photometric data taken from quasars.
Then, a forward selection process is applied, as will be explained in more detail in the next sections.
This is done to build a tree of best performing feature subsets.
This method has to be considered as an alternative to the automatic features extraction used in \citet{2018A&A...609A.111D}.
Both methods can be useful and efficient, depending on the nature of the problem, and on the availability of data and resources.
For this reason, the results obtained with both methods will be compared.
The experimental strategy is based on the application of two different machine learning models and evaluated on the basis of several statistical tools.
In the following these models, kNN and RF, are presented.
The strategy used to perform the feature selection is then depicted in detail and we give a description of the statistical framework used for the experiments' evaluation and of the cross validation algorithm.

\subsection{Regression models}
As mentioned above, our method makes use of kNN and RF models, which are described in detail in the following subsections, while the details regarding the deep convolutional mixture density network (DCMDN) used to compare the results with an automatic features extraction based model  can be found in \citet{2018A&A...609A.111D}.
\subsubsection{kNN}
The kNN \citep{fix_51_discriminatory} is a machine learning model used both for regression and classification tasks \citep{2013AJ....146...22Z}.
This model explores the feature space by estimating the $k$ nearest  points (or neighbours) belonging to the training sample with respect to each test item.
In our case the distance involved is calculated through a Euclidean metric. 
In the case of a regression problem (like redshift estimation), the kNN algorithm is used to find a continuous variable averaging the distances of the k selected neighbours.
The efficiency of the algorithm is strongly related to the choice of the parameter $k$, which represents the number of neighbours to be selected from the training set.
The best choice of this parameter is directly related to the input data, their complexity, and the way in which the input space is sampled.
Clearly, the most simple case is a model with $k=1$.
In this case, a prediction equal to the target of the closest pattern in the training set is associated to each pattern.
Increasing the $k$ parameter could improve the precision of the model (this is due to the increasing generalization capability), but can also generate overfitting \citep{Duda:2000:PC:954544}.
In our experiments, the choice of the $k$ parameter was part of the learning task by evaluating a set of possible values.
The kNN is one of the simplest machine learning algorithms, but even if it could be outperformed by more complex models, it has the advantage of being very fast and in any case quite efficient.
Another possible problem concerning the use of the kNN model is given by possible differences in the range of the input features.
This could generate problems and misleading results in the estimation of distances in the parameter space.
For this reason, all the features used in this work have been normalized using the min-max normalization technique \citep{Aksoy00featurenormalization}.

\subsubsection{Random forest}
The RF \citep{cart84} is one of the most popular ensemble-based machine learning models, and could be used for regression and classification tasks \citep[see][for an application to photometric redshift estimation]{2010ApJ...712..511C}.
It is an ensemble of decision trees, where each tree is meant to partition the feature space in order to find the best split that minimizes the variance.
Each decision tree is built by adding leaf nodes where the input data are partitioned with respect to a different chosen feature, repeating the process for all the possible choices of variables to be split.
In case of a regression problem, the root mean square error (RMSE) is computed for each possible partition, and the partition which minimizes the RMSE is chosen.
The RF averages the results provided by many decision trees, each trained on a different part of the training set through the bagging technique \citep{Breiman1996}.
This avoids overfitting due to single decision trees growing too deep.
Moreover, the decision tree makes use of the bootstrapping technique \citep{Breiman1996} in order to increase the performance and stability of the method and reduce overfitting at the same time.
This consists in giving, as input, a different random sub-sample of the training data to each decision tree.
The RF uses the feature bagging during the training phase.
This consists in selecting a random subset of features at each split.
Bootstrapping and bagging help to avoid correlations between single decision trees, which could appear when training them on the same training set and in the presence of strong features selected multiple times.

\subsection{Features selection strategy}\label{sec:strategy}
The huge number of features evaluated, as described in Sec.~\ref{data}, imposes the need to establish an efficient feature selection process.
In fact, in order to estimate a subset of the best $f=10$ features\footnote{The reason for selecting 10 features is discussed in Sec.~\ref{sec:exp79} and Fig.~\ref{Fig:rmsTrend}}, starting with $r=4,520$ features, would imply, if we want to test all the possible combinations, the following number of experiments:

\begin{equation}
n = \frac{r!}{f!*(r-f)!}=9.7 \times 10^{29}.
\end{equation}

\begin{figure}
   \centering
   \includegraphics[width= 1.0 \columnwidth]{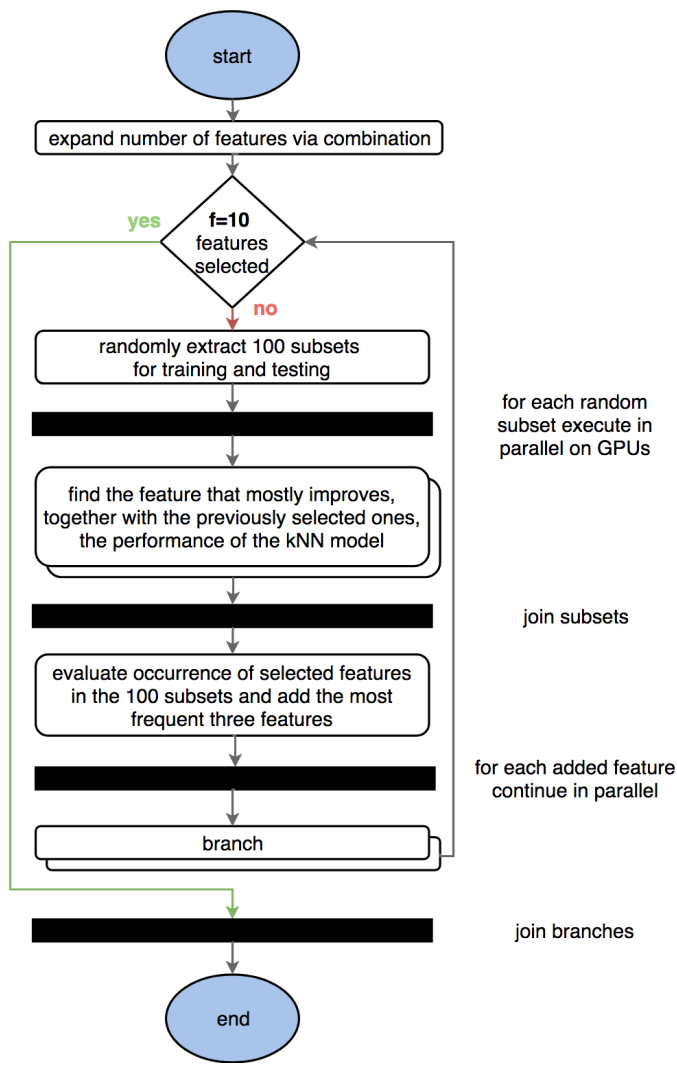}

   \caption{Workflow used to generate tree structure. The black boxes represent states where multiple operations are started in parallel or parallel operations are joined. The iteration is stopped when each branch of the tree has a depth of 10. A five-fold cross validation is applied for every model evaluation step.}\label{Fig:workflow}
\end{figure}

\noindent
Assuming that a nonillion experiments are too many to be performed, a more efficient approach had to be chosen.
Therefore, we decided to apply a forward selection process \citep{Mao2004629} as described in the following.
The number of features used for the experiment was iteratively increased.
In other words, to select the first best feature a kNN model for each of the $r=4,520$ features was trained in a one-dimensional feature space.
Due to the memory limitations of the hardware architecture used, the feature selection was done by performing $100$ kNN experiments, selecting for each of them a random subset of $20,000$ data points and using a five-fold cross validation (see Sec.~\ref{sec:cv} for more details).
The repeated experiments on different training samples were meant to generate statistics of the features in order to identify the most frequently selected ones.
This was done to minimize the biases introduced by the random extraction of the training data.
Since $100$ runs were performed, sometimes more than one feature was selected.
The basic idea behind the proposed strategy is to select a limited number of best-performing features per step.
The number of features which were actually selected were chosen by evaluating the occurrence of each of them as the best feature in all of the $100$ runs.
Therefore, for each iteration a minimum of one and a maximum of three features were selected.
After choosing the best features, they were fixed and the next run was performed in order to choose the subsequent features.
This method was iterated until the tenth feature was selected.
A tree with a maximum branching number of three was derived, because in every step a maximum number of three features that best improve the model were chosen.
Each branch can be seen as a set of best-performing-feature combinations.
The necessity of performing a high number of experiments on different data subsets is caused by the slightly varying behaviour of the kNN model with respect to different input patterns.
The whole workflow is summarized in Fig.~\ref{Fig:workflow}.
The cross validation, moreover, was used in order to further reduce any risk of overfitting.

\subsection{GPU parallelization for kNN}
The feature selection is done by parallelizing the experiments on a GPU cluster.
The massive use of GPUs proved to be mandatory in order to deal with such an amount of data, features, $k$ values, and runs on randomly sampled data-sets.
Following \citet{heinermann2013} and \citet{DBLP:conf_esann_GiesekePOI14}, the kNN algorithm has been parallelized by using GPUs.
Typically, GPU-based programs are composed by a host program running on central processing unit (CPU) and a kernel program running on the GPU itself, which is parallelized on the GPU cores in several threads or kernel instances.
This scheme is particularly adapted to kNN models, due to the advantages obtained by parallelizing matrix multiplications.
In the code used for this work \citep{DBLP:conf_esann_GiesekePOI14} the calculation is performed by generating matrices containing the distances of the selected features from the query object.
This calculation is entirely performed on the GPU, while the CPU is mainly used for synchronization and for updating a vector containing the selected features at every step.
The approach based on this method proved to speed up the calculation by a factor of $\sim 150$.
We modified the given code to start the selection process with a given set of already selected features.
This was done to enable the generation of the feature trees based on $100$ random subsets.

\subsection{Statistical estimators and cross validation}\label{sec:cv}
The results have been evaluated using the following set of statistical scores for the quantity $\Delta z = (\zs-\zp)/(1+\zs)$ expressing the estimation error\footnote{We note that $\Delta z$ denotes the normalized error in redshift estimation and not the usually used plain error.} on the objects in the blind validation set:

\begin{itemize}
\item{bias: defined as the mean value of the normalized residuals $\Delta z$;}
\item{RMSE: root mean square error;}
\item{NMAD: normalized median absolute deviation of the normalized residuals, defined as $NMAD(\Delta z) = 1.48 \times median(\left|\Delta z_i - median(\Delta z)\right|)$;}
\item{CRPS: the continuous rank probability score \citep{2000WtFor..15..559H} is a proper score to estimate how well a single value is represented by a distribution. It is used following \citet{2018A&A...609A.111D}.}
\end{itemize}

\noindent
The prediction of redshifts in a probabilistic framework has many advantages.
The ability of reporting the uncertainty is the most important one to mention.
In order to correctly evaluate the performance of the features in a probabilistic setting, the CRPS was added to the set of scores.
By using the RF as a quantile regression forest and fitting a mixture of Gaussians to the predictions of the ensemble members, a probability distribution can be generated and the CRPS can be calculated.
The DCMDN, by definition, predicts density distributions that are represented by their mean when calculating the scores used for point estimates.

As stated before, all the indicators are then averaged on the $k$ folds of the cross validation.
Through this approach, the standard deviation is also obtained as a measure of the error on each statistical estimator.
We do not report those values as the errors were small enough to be considered negligible.
Cross validation \citep{Kohavi95astudy} is a statistical tool used to estimate the generalization error.
The phenomenon of overfitting arises when the model is too well adapted to the training data.
In this case, the performance on the test set will be poor as the model is not general enough.
A validation set is defined, in order to test this generalization of the model, with respect to the training data, on an unseen and omitted set of data.
In particular, cross validation becomes necessary when dealing with small training sets or high-dimensional feature spaces.

In this kind of approach, the data-set is divided into $k$ subsets and each of them is used for the prediction phase, while all the $k-1$ subsets constitute the training set.
The training is then repeated $k$ times, using all the subsets.
The final performance is obtained by averaging the results of the single folds and the error on the performance is obtained by evaluating the standard deviation of the results coming from the different folds. 
In this work, we adopt a k-fold cross validation approach, with $k=5$ for the kNN experiments and $k=10$ for the RF experiments. 

\begin{figure}[!t]
   \centering
   \includegraphics[width= 1.0 \columnwidth]{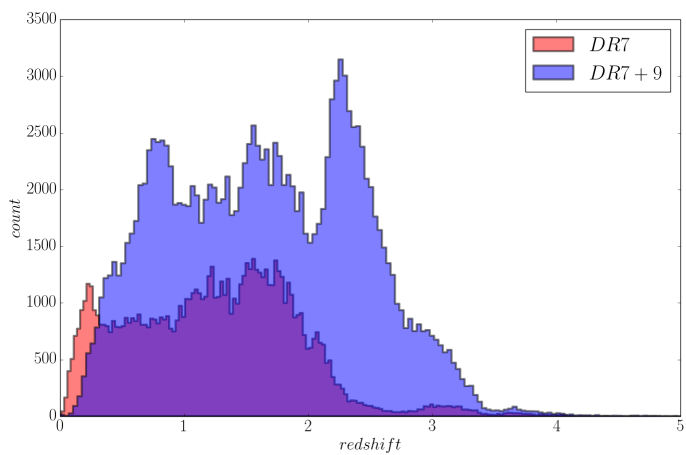}
   \caption{Histogram showing the redshift distribution of the catalogues with objects from DR7 only and DR7 plus DR9. The distribution for the catalogue DR7b is not reported here because the difference with respect to catalogue DR7a is practically negligible.}\label{Fig:histoz}
\end{figure}

\section{Data}\label{data}
In the following subsections the details about the data-set used and the feature combinations performed for the experiments are outlined.

\subsection{Data-sets}
The experiments are based on quasar data extracted from the SDSS DR7 (\citealt{2009ApJS..182..543A}) and SDSS DR9 (\citealt{2012ApJS..203...21A}).
Three catalogues have been retrieved for the experiments.
Moreover, images for the DCMDN experiments have been downloaded making use of Hierarchical Progressive Survey \citep[HiPS][]{2015A&A...578A.114F}.

\subsubsection*{Catalogue DR7a}
Catalogue DR7a is the most conservative with respect to the presence of bad data or problematic objects.
It is based on DR7 only, with clean photometry and no missing data; the query used is reported in Appendix \ref{queryQSO}.
Furthermore, to be more conservative, we checked the spectroscopic redshifts in two different data releases (9 and 12) and we decided to cut all the objects with a discrepancy in {\zs} not fulfilling the given criteria\\[5pt]

\begin{tabular}{ll}
$|z_{DR7\phantom{1}}  - z_{DR9\phantom{1}} |$ < 0.01, \textrm{and} \\
$|z_{DR7\phantom{1}}  - z_{DR12}|$ < 0.01, \textrm{and} \\
$|z_{DR12} - z_{DR9\phantom{1}} |$ < 0.01.
\end{tabular}
\\[5pt]

\noindent
The final catalogue contains $83,982$  objects with a spectroscopically determined redshift.

\subsubsection*{Catalogue DR7b}
Catalogue DR7b has been obtained using the same query used for Catalogue DR7a, but removing the image processing flags.
This has been done in order to verify if the presence of objects previously discarded by the use of these flags could affect the feature selection process.
The catalogue has been cleaned by removing all the objects with NaNs and errors bigger than a value of one, ending with a catalogue containing $97,041$ objects.

\subsubsection*{Catalogue DR7+9}
Catalogue DR7+9 has been prepared mixing quasars from DR7 and DR9 in order to perform the feature selection with a different and more complete redshift distribution.
The difference in the redshift distribution of the two catalogues can be seen from the histogram in Fig.~\ref{Fig:histoz}.
The catalogue has been cleaned with the same procedure adopted for Catalogue DR7b and the common objects between DR7 and DR9 have been used only once.
This produced a catalogue of $152,137$ objects.
In the following sections, the results obtained with this catalogue are discussed in depth.

\subsection{Classic features}

In classic redshift estimation experiments for quasars and galaxies, as can be found in the literature \citep[e.g.][]{2007ApJ...663..752D}, for SDSS data colours are mainly used as features.
To be comparable, we decided to use a set of ten features as our benchmark feature set.
Colours of the adjacent filterbands for the point spread function (PSF) and model magnitudes are used together with the plain PSF and model magnitudes.
In SDSS, the model magnitudes are the best fitting result of an exponential or de Vaucouleurs model.
All {\sffamily{Classic$_{10}$}} features can be found in the first column of Table~\ref{tab:best_features}.

\subsection{Combined features}

\begin{table}[!t]\resizebox{.49\textwidth}{!}{
\centering\begin{tabular}{lllll}
                        &Magnitudes             & $\sigma$              & Radii           & Ellipticities \\ \hline
                        &modelMag / Extinction          & \checkmark            & devRad          & devAB         \\
                        &petroMag / Extinction          & \checkmark            & expRad          & expAB         \\
                        &psfMag / Extinction            & \checkmark            & petroRad                &               \\
                        &devMag / Extinction            & \checkmark            & petroR50                &               \\
                        &expMag / Extinction            & \checkmark            & petroR90                &               \\\hline\hline
Plain   & 25 + 25 dereddened            & \hspace{1ex}25                        & \hspace{1ex}25                  & \hspace{1ex}10                \\      \hline
Combined        & $1,225$ Differences   & 300 Pairs      & 300 Differences       & \hspace{1ex}45 Differences      \\
                        &  $2,450$ Ratios               &                       &                 & \hspace{1ex}90 Ratios \\\hline\hline
Total $4,520$   & $3,725$                       & 325                   & 325                     & 145           \\ \hline
\end{tabular}
}
\caption{Types of features downloaded from SDSS and their combinations in order to obtain the final catalogue used for the experiments. The number of each feature type is given alongside with the final number of synthetically derived features.}\label{data_and_features}
\end{table}

For each of the three catalogues, the features concerning magnitudes and their errors, radii, ellipticities, and extinction are retrieved.
An overview of the features is shown in Table~\ref{data_and_features}.
Magnitudes that have been corrected for extinction are denoted with an underline indicating that, for example, \ul{$u_{model}$} is equivalent to $u_{model} - u_{extinction}$.
The parameter space has been enriched by performing several combinations of the original features \citep{DBLP:conf_esann_GiesekePOI14}. 
A similar feature generation approach was applied also in \citet{2014ASPC..485..425P} but with a limited set of plain features and combination rules.
In other words, the magnitude features were combined obtaining all the pairwise differences and ratios, both in the normal and dereddened version.
The errors on the magnitudes have been composed taking their quadratic sums.
Finally, radii and ellipticities have been composed through pairwise differences with ratios only for the ellipticities.
The final catalogue consists of $4,520$ features for each data item.
It has to be noted that the {\sffamily{Classic$_{10}$}} features are of course included in this set of features.
In Table~\ref{data_and_features}, the types and amounts of the features obtained following this strategy are specified.
As appears from the table, the feature combinations can be divided into several groups:
\begin{itemize}
    \item{simple features: magnitudes, radii, and ellipticities as downloaded from the SDSS database.}
    \item {differences: pairwise differences of the simple features; colour indexes are a subset of this group utilizing only adjacent filters.}
    \item{ratios: ratios between the simple features; an important subset of this group is the one containing ratios between different magnitudes of the same filter; we will define this subset as photometric ratios.}
    \item{errors: errors on the simple features and their propagated compositions.}
\end{itemize}
As we will see in the following, the ratios group, and its subgroup, the photometric ratios, are particularly important for the redshift estimation experiments.

\section{Experiments and results}\label{results}
The feature selection was performed applying the forward selection strategy, as described in Sec.~\ref{sec:strategy}, on the three catalogues.
The verification of the resulting feature sets was performed using the RF.
This algorithm is widely used in literature, and therefore the results obtained here can be easily compared to those with different feature selection strategies.

In addition, experiments using the classic features were performed, in order to compare their performances with the proposed selected features.
Already at an early stage of the experiments, it turned out that only four selected features are sufficient to achieve a performance comparable to classic features.
Therefore the scores are always calculated separately for the full set of ten selected features ({\sffamily{Best$_{10}$}}) and the first four ({\sffamily{Best$_{4}$}}) features only.
To compare the results with a fully automated feature extraction and feature selection approach, a DCMDN was also used for the experiments.

It has to be noted that in some cases the same features sets have been found but exhibiting a different ordering.
In these cases, all the subsets have been kept for the sake of correctness.
In the next subsections the three experiments and the corresponding results are shown.
The two experiments with Catalogue DR7a and DR7b are designed to provide results that are comparable to the literature.
For a scientifically more interesting interpretation, the less biased, not flagged, and more representative Catalogue DR7+9 was used for the main experiment.
Therefore, only the results and performances of the first two experiments are given in a summarized representation, reserving more space for a detailed description of Experiment DR7+9.
Further details concerning the results obtained with Catalogue DR7a and DR7b are shown in Appendix~\ref{Sec:additions}.

\begin{table}[!ht]\resizebox{1.0\columnwidth}{!}{
\centering\begin{tabular}{c|c|c|c}
{\sffamily{Classic$_{10}$}} & DR7a Best$_{10}$ & DR7b Best$_{10}$ & DR7+9 Best$_{10}$\\ 
\hline
$r_{psf}$ & $i_{psf}/i_{model}$ & $i_{psf}/i_{model}$ & $i_{petro}/i_{psf}$\\
$r_{model}$ & $\text{\ul{$g_{psf}$}}/\text{\ul{$u_{model}$}}$ & $\text{\ul{$g_{psf}$}}/\text{\ul{$u_{model}$}}$ & $\text{\ul{$g_{psf}$}}-\text{\ul{$u_{model}$}}$\\
$u_{psf}-g_{psf}$ & $\text{\ul{$r_{psf}$}}/\text{\ul{$i_{model}$}}$ & $\text{\ul{$r_{psf}$}}/\text{\ul{$i_{model}$}}$ & $\text{\ul{$i_{exp}$}}/\text{\ul{$r_{psf}$}}$\\
$g_{psf}-r_{psf}$ & $\text{\ul{$i_{dev}$}}/\text{\ul{$i_{psf}$}}$ & $\text{\ul{$i_{dev}$}}/\text{\ul{$i_{psf}$}}$ & $\sqrt{\sigma_{r_{model}}^2+\sigma_{r_{dev}}^2}$\\
$r_{psf}-i_{psf}$ & $\text{\ul{$r_{psf}$}}/\text{\ul{$g_{model}$}}$ & $\text{\ul{$z_{psf}$}}/\text{\ul{$i_{model}$}}$ & $\text{\ul{$r_{psf}$}}/\text{\ul{$g_{exp}$}}$\\
$i_{psf}-z_{psf}$ & $\text{\ul{$i_{psf}$}}/\text{\ul{$z_{model}$}}$ & $\text{\ul{$r_{psf}$}}/\text{\ul{$g_{exp}$}}$ & $\text{\ul{$i_{psf}$}}/\text{\ul{$z_{model}$}}$\\
$u_{model}-g_{model}$ & $r_{psf}-r_{petro}$ & $r_{psf}-r_{petro}$ & $i_{psf}-i_{dev}$\\
$g_{model}-r_{model}$ & $\sqrt{\sigma_{r_{model}}^2+\sigma_{g_{exp}}^2}$ & $i_{psf}-i_{petro}$ & $r_{petro}/r_{psf}$\\
$r_{model}-i_{model}$ & $z_{model}/z_{psf}$ & $\text{\ul{$z_{model}$}}/\text{\ul{$z_{psf}$}}$ & $\text{\ul{$i_{psf}$}}-\text{\ul{$r_{model}$}}$\\
$i_{model}-z_{model}$ & $i_{psf}-i_{petro}$ & $\sqrt{\sigma_{g_{model}}^2+\sigma_{g_{dev}}^2}$ & $z_{exp}/z_{psf}$\\
\hline
\end{tabular}
}
\caption{Classic and best feature subsets obtained by the feature selection process of the experiments on the three catalogues. After the selection process, the RF was used to identify the feature branches of the corresponding trees that show the best performance.}\label{tab:best_features}
\end{table}

\begin{table}\resizebox{1.0\columnwidth}{!}{
\centering
\begin{tabular}{|l|l|r|r|r|r|}
\hline
 \textbf{Exp} & \textbf{Set} & \textbf{\# Features}  & \textbf{mean}  & \textbf{RMSE}  & \textbf{NMAD} \\
\hline 
\multirow{4}{*}{\textbf{DR7a}} & {\sffamily{Classic$_{10}$}}  & 10 & -0.024 & 0.163 & 0.051 \\
&{\sffamily{Best$_{4}$}} & 4 & -0.023 & 0.163 & 0.080 \\
&{\sffamily{Best$_{10}$}} & 10 & -0.014 & 0.124 & 0.044 \\
&\sffamily{DCMDN} & 65,536 & -0.020 & 0.145 & 0.043 \\
\hline
\hline 
\multirow{4}{*}{\textbf{DR7b}}&{\sffamily{Classic$_{10}$}}  & 10 & -0.030 & 0.180 & 0.059 \\
&{\sffamily{Best$_{4}$}} & 4 & -0.027 & 0.183 & 0.087 \\
&{\sffamily{Best$_{10}$}} & 10 & -0.019 & 0.145 & 0.050 \\
&\sffamily{DCMDN} & 65,536 & -0.024 & 0.171 & 0.032 \\
\hline
\hline 
\multirow{4}{*}{\textbf{DR7+9}}&{\sffamily{Classic$_{10}$}}  & 10 & -0.033 & 0.207 & 0.073 \\
&{\sffamily{Best$_{4}$}} & 4 & -0.032 & 0.206 & 0.100 \\
&{\sffamily{Best$_{10}$}} & 10 & -0.023 & 0.174 & 0.060 \\
&\sffamily{DCMDN} & 65,536 & -0.027 & 0.184 & 0.037 \\
\hline\end{tabular}
}
\caption{Summary of the scores obtained with the RF and DCMDN models in the three experiments. The DCMDN automatically extracted $65,536$ features for each experiment. The resulting scores are also given.}\label{tab:results_summary}
\end{table}

\begin{table}\footnotesize
\centering
\begin{tabular}{|l|r|}
\hline
\textbf{DR7a} & \textbf{CRPS}\\
\hline 
{\sffamily{Classic$_{10}$}} & 0.110\\
{\sffamily{Best$_{4}$}}  & 0.154 \\
{\sffamily{Best$_{10}$}} & 0.089\\
\sffamily{DCMDN} & 0.099\\
\hline\end{tabular}
\begin{tabular}{|l|r|}
\hline
\textbf{DR7b} & \textbf{CRPS}\\
\hline 
{\sffamily{Classic$_{10}$}} & 0.131\\
{\sffamily{Best$_{4}$}}  & 0.172 \\
{\sffamily{Best$_{10}$}} & 0.106\\
\sffamily{DCMDN} & 0.124\\
\hline\end{tabular}
\begin{tabular}{|l|r|}
\hline
\textbf{DR7+9} & \textbf{CRPS}\\
\hline 
{\sffamily{Classic$_{10}$}} & 0.167\\
{\sffamily{Best$_{4}$}}  & 0.203 \\
{\sffamily{Best$_{10}$}} & 0.140\\
\sffamily{DCMDN} & 0.146\\
\hline\end{tabular}
\caption{Table showing the performance of the different feature subsets with respect to the CRPS score for the three catalogues.}\label{tab:crps_results}
\end{table}

\subsection{Experiment DR7a}
The feature selection on the Catalogue DR7a produced $22$ subsets of ten features each.
Only $20$ features, of the initial $4,520$, compose the tree.
The three features,
\begin{itemize}
    \item {$\text{\ul{$g_{psf}$}}/\text{\ul{$u_{model}$}}$}
    \item {$\text{\ul{$i_{psf}$}}/\text{\ul{$z_{model}$}}$}
    \item {$z_{model}/z_{psf,}$}
\end{itemize}

\noindent appear in all the possible branches.
For all presented feature sets, the RF experiments were performed.
The best performing ten features are indicated in the second column of Table~\ref{tab:best_features} (DR7a subset) in the order of their occurrence.
The performances are compared with the results of the {\sffamily{Classic$_{10}$}} features presented in the first column of the same table.
A summary of the most important results is shown in the first section of Table~\ref{tab:results_summary}.
As shown in Table~\ref{tab:results_summary}, the experiment with the {{\sffamily{Best$_{10}$}}} subset outperforms the experiment with the {\sffamily{Classic$_{10}$}} features with respect to all the statistical scores.

Moreover, in Table~\ref{tab:results_summary} the results obtained using the DCMDN are shown in order to compare the predictions with a model based on automatic features selection.
The DCMDN model automatically extracts $65,536$ features from images in the five filters \emph{ugriz} of size $16 \times 16$ $pixel^{2}$.
This model is meant to generate probability density functions (PDFs) in the form of Gaussian mixtures instead of point estimates.
Therefore, in order to calculate the scores, the weighted mean of every PDF with respect to the mixture components has been estimated.
As shown in the table, the performance is superior with respect to the {\sffamily{Classic$_{10}$}} features and the {\sffamily{Best$_{4}$}} subset, but it is outperformed by the {\sffamily{Best$_{10}$}} subset of features.
The performances of these four sets have been compared using the CRPS score, as reported in the left section of Table~\ref{tab:crps_results}.
Those results are consistent with the previously found results.
A detailed listing of the results is given in Appendix~\ref{Sec:additions} with the individual feature tree being visualized as a chord diagram \citep{Krzywinski18062009}.

\subsection{Experiment DR7b}
In the experiment performed with Catalogue DR7b, the proposed model selected $26$ features generating $41$ subset combinations.
Only the following two features appear in all the subsets:

\begin{itemize}
    \item {$i_{psf}/i_{petro}$}
    \item {$\text{\ul{$g_{psf}$}}/\text{\ul{$u_{model}$}}$.}
\end{itemize}

\noindent
From the RF validation runs, the subset reported in the third column of Table~\ref{tab:best_features} (DR7b) produces the best performance.
The most important results are shown in the second section of Table~\ref{tab:results_summary}, in which the results obtained with the previous experiment (DR7a) are confirmed.
This is valid considering both the RMSE and the CRPS indicators. 
The CRPS is shown in the middle section of Table~\ref{tab:crps_results}.
Therefore, the performance given using the {\sffamily{Best$_{10}$}} subset is superior to that using the {\sffamily{Classic$_{10}$}} features.
The DCMDN model is outperformed too.
Several features can be found in both experiments with catalogues DR7a and DR7b and the general structure of the tree between the two experiments is comparable.
Therefore, the exclusion of photometric flags seems not to affect substantially the global process of feature selection.
It can be noticed, however, that the general performance degrades.
This is due to the increased presence of objects characterized by a less clean photometry.
The detailed feature selection results for this experiment and the chord diagram are also shown in Appendix~\ref{Sec:additions}.

\subsection{Experiment DR7+9}\label{sec:exp79}

\begin{table*}[p]\resizebox{1.0\textwidth}{!}{
\centering
\definecolor{colora}{RGB}{230,46,46}
\definecolor{colorb}{RGB}{230,116,46}
\definecolor{colorc}{RGB}{230,187,46}
\definecolor{colord}{RGB}{202,230,46}
\definecolor{colore}{RGB}{132,230,46}
\definecolor{colorf}{RGB}{61,230,46}
\definecolor{colorg}{RGB}{46,230,101}
\definecolor{colorh}{RGB}{46,230,171}
\definecolor{colori}{RGB}{46,217,230}
\definecolor{colorj}{RGB}{46,147,230}
\definecolor{colork}{RGB}{46,77,230}
\definecolor{colorl}{RGB}{86,46,230}
\definecolor{colorm}{RGB}{156,46,230}
\definecolor{colorn}{RGB}{230,46,230}

\begin{tikzpicture}
\matrix (m)[matrix of nodes, style={nodes={rectangle,minimum width=8em,  row sep=-\pgflinewidth, column sep=-\pgflinewidth}}, ampersand replacement =\&]
{
  {\bf id} \&
  {\bf Feature 1} \& {\bf Feature 2} \& {\bf  Feature 3} \&  {\bf Feature 4} \&  {\bf Feature 5} \&  {\bf Feature 6} \&  {\bf Feature 7} \&  {\bf Feature 8} \& {\bf  Feature 9} \& {\bf  Feature 10} \\
\hline
$1$     \&      \raisebox{1.2em}{$$}    \&      \raisebox{1.2em}{$$}    \&      \raisebox{1.2em}{$$}    \&      \raisebox{1.2em}{$$}    \&      \raisebox{1.2em}{$$}    \&      \raisebox{1.2em}{$$}    \&      \raisebox{1.2em}{$$}    \&      \raisebox{1.2em}{$$}    \&      \raisebox{1.2em}{$$}\colorbox{white}{${\text{\ul{$i_{psf}$}}-\text{\ul{$i_{dev}$}}}$}   \&      \raisebox{1.2em}{$$}\raisebox{-1.2em}{$$}\colorbox{white}{${z_{psf}-z_{model}}$}        \\
$2*$    \&      $ $     \&      $ $     \&      $ $     \&      $ $     \&      $ $       \&      $ $     \&      $ $     \&      \colorbox{white}{${r_{petro}/r_{psf}}$} \&      \raisebox{-1.4em}{$$}   \&      \raisebox{1.2em}{$$}\colorbox{white}{${z_{exp}/z_{psf}}$}       \\
$3$     \&      $ $     \&      $ $     \&      $ $     \&      \colorbox{white}{${\sqrt{\sigma_{r_{model}}^{2}+\sigma_{r_{dev}}^{2}}}$}        \&      $ $       \&      $ $     \&      $ $     \&      $ $     \&      \raisebox{1.4em}{$$}\colorbox{white}{${\text{\ul{$i_{psf}$}}-\text{\ul{$r_{model}$}}}$} \&      \raisebox{-1.2em}{$$}   \\
$4$     \&      $ $     \&      $ $     \&      $ $     \&      $ $     \&      $ $       \&      $ $     \&      $ $     \&      \raisebox{-1.2em}{$$}   \&      \raisebox{-1.2em}{$$}   \&      \raisebox{1.2em}{$$}\colorbox{white}{${\text{\ul{$i_{psf}$}}-\text{\ul{$i_{dev}$}}}$}   \\
$5$     \&      \colorbox{white}{${i_{petro}/i_{psf}}$} \&      \colorbox{white}{${\text{\ul{$g_{psf}$}}-\text{\ul{$u_{model}$}}}$}     \&      \colorbox{white}{${\text{\ul{$i_{exp}$}}/\text{\ul{$r_{psf}$}}}$}       \&      \raisebox{-1.2em}{$$}   \&      \colorbox{white}{${\text{\ul{$r_{psf}$}}/\text{\ul{$g_{exp}$}}}$}       \&      \colorbox{white}{${\text{\ul{$i_{psf}$}}/\text{\ul{$z_{model}$}}}$}     \&      \colorbox{white}{${i_{psf}-i_{dev}}$}   \&      \raisebox{1.2em}{$$}\raisebox{-1.2em}{$$}\colorbox{white}{${z_{psf}-z_{model}}$}        \&      \raisebox{1.2em}{$$}\raisebox{-1.2em}{$$}\colorbox{white}{${r_{petro}/r_{psf}}$}        \&      \raisebox{-1.2em}{$$}   \\
$6$     \&      $ $     \&      $ $     \&      $ $     \&      \raisebox{1.2em}{$$}    \&      $ $       \&      $ $     \&      $ $     \&      \raisebox{1.2em}{$$}\colorbox{white}{${r_{petro}/r_{psf}}$}     \&      \raisebox{1.2em}{$$}\raisebox{-1.4em}{$$}\colorbox{white}{${\text{\ul{$i_{psf}$}}-\text{\ul{$i_{dev}$}}}$}      \&      \raisebox{1.2em}{$$}\raisebox{-1.4em}{$$}\colorbox{white}{${z_{psf}-z_{model}}$}        \\
$7$     \&      $ $     \&      $ $     \&      $ $     \&      \colorbox{white}{${\sqrt{\sigma_{g_{model}}^{2}+\sigma_{r_{dev}}^{2}}}$}        \&      $ $       \&      $ $     \&      $ $     \&      \raisebox{-1.2em}{$$}   \&      \raisebox{1.4em}{$$}\raisebox{-1.2em}{$$}\colorbox{white}{${z_{psf}-z_{model}}$}        \&      \raisebox{1.4em}{$$}    \\
$8$     \&      $ $     \&      $ $     \&      $ $     \&      $ $     \&      $ $       \&      $ $     \&      $ $     \&      \raisebox{1.2em}{$$}\colorbox{white}{${z_{psf}-z_{model}}$}     \&      \raisebox{1.2em}{$$}\raisebox{-1.2em}{$$}\colorbox{white}{${r_{petro}/r_{psf}}$}        \&      \colorbox{white}{${\text{\ul{$i_{psf}$}}-\text{\ul{$i_{dev}$}}}$}       \\
$9$     \&      \raisebox{-1.2em}{$$}   \&      \raisebox{-1.2em}{$$}   \&      \raisebox{-1.2em}{$$}   \&      \raisebox{-1.2em}{$$}   \&      \raisebox{-1.2em}{$$}   \&      \raisebox{-1.2em}{$$}   \&      \raisebox{-1.2em}{$$}   \&      \raisebox{-1.2em}{$$}   \&      \raisebox{1.2em}{$$}\raisebox{-1.2em}{$$}\colorbox{white}{${r_{psf}-r_{petro}}$}        \&      \raisebox{-1.2em}{$$}   \\
};

\begin{pgfonlayer}{myback}
\fillpattern[dots][colora]{m-2-2}{m-10-2}
\fillpattern[horizontal lines][colorb]{m-2-3}{m-10-3}
\fillpattern[vertical lines][colorc]{m-2-4}{m-10-4}
\fillpattern[north west lines][colord]{m-2-5}{m-6-5}
\fillpattern[north west lines][colore]{m-7-5}{m-10-5}
\fillpattern[vertical lines][colorf]{m-2-6}{m-10-6}
\fillpattern[vertical lines][colorg]{m-2-7}{m-10-7}
\fillpattern[horizontal lines][colorh]{m-2-8}{m-10-8}
\fillpattern[dots][colori]{m-2-9}{m-5-9}
\fillpattern[horizontal lines][colorj]{m-6-9}{m-6-9}
\fillpattern[dots][colori]{m-7-9}{m-8-9}
\fillpattern[horizontal lines][colorj]{m-9-9}{m-10-9}
\fillpattern[horizontal lines][colorn]{m-2-10}{m-3-10}
\fillpattern[horizontal lines][colork]{m-4-10}{m-5-10}
\fillpattern[dots][colori]{m-6-10}{m-6-10}
\fillpattern[horizontal lines][colorn]{m-7-10}{m-7-10}
\fillpattern[horizontal lines][colorj]{m-8-10}{m-8-10}
\fillpattern[dots][colori]{m-9-10}{m-9-10}
\fillpattern[horizontal lines][colorl]{m-10-10}{m-10-10}
\fillpattern[horizontal lines][colorj]{m-2-11}{m-2-11}
\fillpattern[dots][colorm]{m-3-11}{m-4-11}
\fillpattern[horizontal lines][colorn]{m-5-11}{m-6-11}
\fillpattern[horizontal lines][colorj]{m-7-11}{m-7-11}
\fillpattern[horizontal lines][colorn]{m-8-11}{m-10-11}
\end{pgfonlayer}
\end{tikzpicture}
}
\caption{Detailed feature branches obtained from the feature selection for the DR7+9 experiment. The $2nd$ branch, indicated with the $*$ symbol, is the best performing subset with respect to the experiments using the RF. The ratios and photometric ratios are indicated, respectively, with vertical lines and dots. The differences are with horizontal lines and the errors are with north west lines. The colour code for the features is the same as shown in the chord diagram in Fig.~\ref{Fig:chorddr79}.}\label{tab:treedefinition_dr79}\end{table*}

\begin{figure*}[p]
\definecolor{colora}{RGB}{230,46,46}
\definecolor{colorb}{RGB}{230,116,46}
\definecolor{colorc}{RGB}{230,187,46}
\definecolor{colord}{RGB}{202,230,46}
\definecolor{colore}{RGB}{132,230,46}
\definecolor{colorf}{RGB}{61,230,46}
\definecolor{colorg}{RGB}{46,230,101}
\definecolor{colorh}{RGB}{46,230,171}
\definecolor{colori}{RGB}{46,217,230}
\definecolor{colorj}{RGB}{46,147,230}
\definecolor{colork}{RGB}{46,77,230}
\definecolor{colorl}{RGB}{86,46,230}
\definecolor{colorm}{RGB}{156,46,230}
\definecolor{colorn}{RGB}{230,46,230}
\begin{minipage}[]{0.76\textwidth}
\centering
\includegraphics[width=1.0\textwidth]{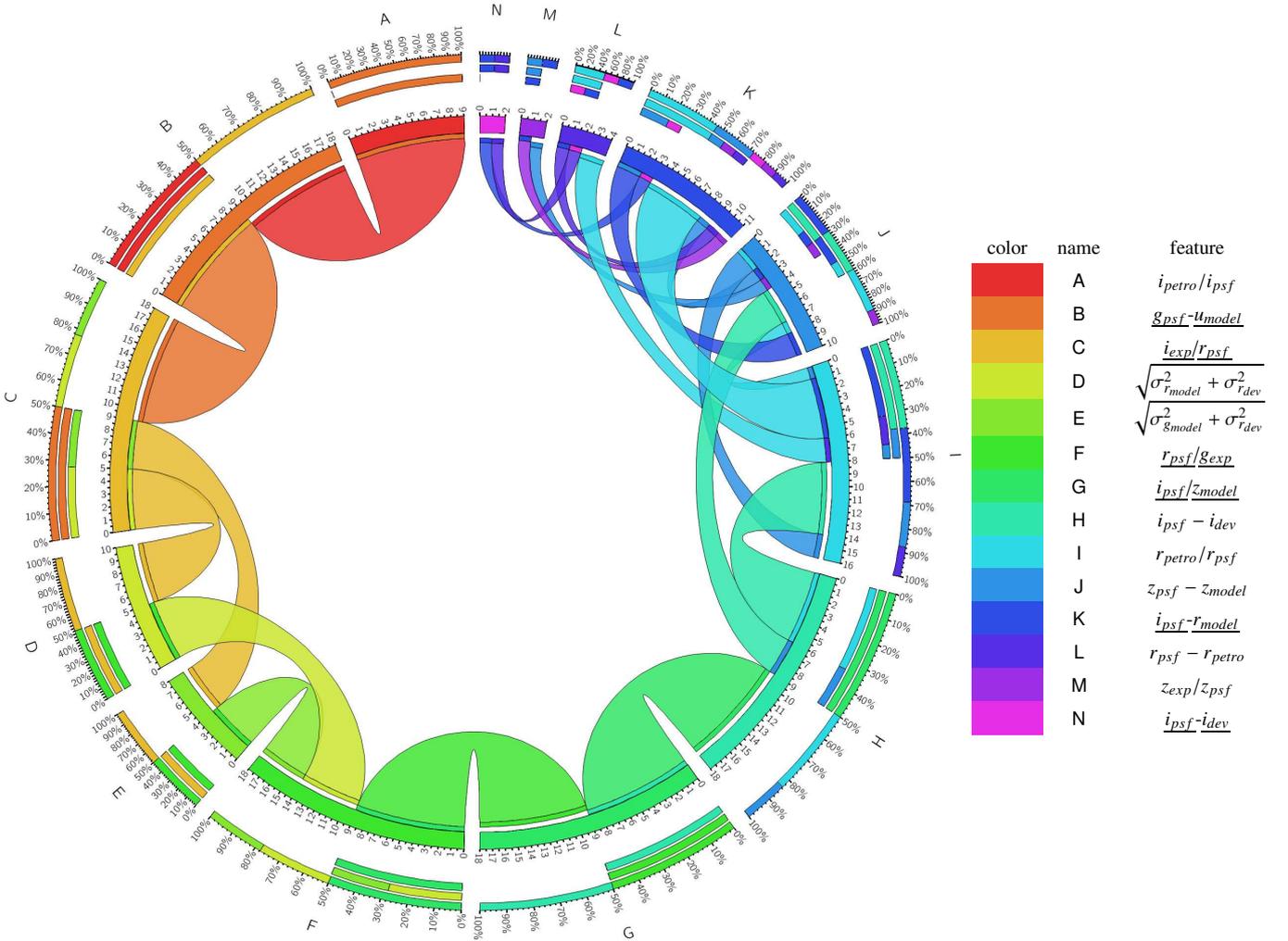}
\end{minipage}\hfill
\begin{minipage}[]{0.235\textwidth}
\renewcommand{\arraystretch}{1.5}
\tiny
\begin{tabular}{ccc}
color & name & feature\\
\colsquare{colora}&{\sffamily A}        &       $i_{petro}/i_{psf}$     \\
\colsquare{colorb}&{\sffamily B}        &       $\text{\ul{$g_{psf}$}-\ul{$u_{model}$}}$        \\
\colsquare{colorc}&{\sffamily C}        &       $\text{\ul{$i_{exp}$}/\ul{$r_{psf}$}}$  \\
\colsquare{colord}&{\sffamily D}        &       $\sqrt{\sigma_{r_{model}}^{2}+\sigma_{r_{dev}}^{2}}$    \\
\colsquare{colore}&{\sffamily E}        &       $\sqrt{\sigma_{g_{model}}^{2}+\sigma_{r_{dev}}^{2}}$    \\
\colsquare{colorf}&{\sffamily F}        &       $\text{\ul{$r_{psf}$}/\ul{$g_{exp}$}}$  \\
\colsquare{colorg}&{\sffamily G}        &       $\text{\ul{$i_{psf}$}/\ul{$z_{model}$}}$        \\
\colsquare{colorh}&{\sffamily H}        &       $i_{psf}-i_{dev}$       \\
\colsquare{colori}&{\sffamily I}        &       $r_{petro}/r_{psf}$     \\
\colsquare{colorj}&{\sffamily J}        &       $z_{psf}-z_{model}$     \\
\colsquare{colork}&{\sffamily K}        &       $\text{\ul{$i_{psf}$}-\ul{$r_{model}$}}$        \\
\colsquare{colorl}&{\sffamily L}        &       $r_{psf}-r_{petro}$             \\
\colsquare{colorm}&{\sffamily M}        &       $z_{exp}/z_{psf}$\\
\colsquare{colorn}&{\sffamily N}        &       $\text{\ul{$i_{psf}$}-\ul{$i_{dev}$}}$  \\
\end{tabular}\end{minipage}
\renewcommand{\arraystretch}{1}
   \caption{Chord diagram of the features derived in Experiment DR7+9. Every feature is associated to a specific colour, and starting from the first feature A it is possible to follow all the possible paths of the tree, depicting the different feature subsets. Ordered from outside to inside, the external arcs represent the occurrences of a particular feature: the total percentage of the individual connections, the numbers and sources of connections entering, and the numbers and targets of connections exiting. (Note the branches splitting in feature C and re-joining in feature F).
   }\label{Fig:chorddr79}
\end{figure*}

The feature selected from the Catalogue DR7+9 are shown in Table~\ref{tab:treedefinition_dr79}. 
In Fig.~\ref{Fig:chorddr79} a chord diagram is given to visualize the structure of the individual subsets.
In this experiment the model selected $14$ individual features grouped in nine subsets.
Due to the different redshift distribution, different features are selected with respect to the previous experiments.
The following six features are in common between all the subsets:

\begin{itemize}
    \item {$i_{psf}-i_{dev}$}
    \item {$\text{\ul{$i_{psf}$}}/\text{\ul{$z_{model}$}}$}
    \item {$\text{\ul{$g_{psf}$}}-\text{\ul{$u_{model}$}}$}
    \item {$i_{petro}/i_{psf}$}
    \item {$\text{\ul{$r_{psf}$}}/\text{\ul{$g_{exp}$}}$}
    \item {$\text{\ul{$i_{exp}$}}/\text{\ul{$r_{psf}$}}$.}
\end{itemize}

\begin{figure*}[!ht]
\centering
\includegraphics[width=1.0\textwidth]{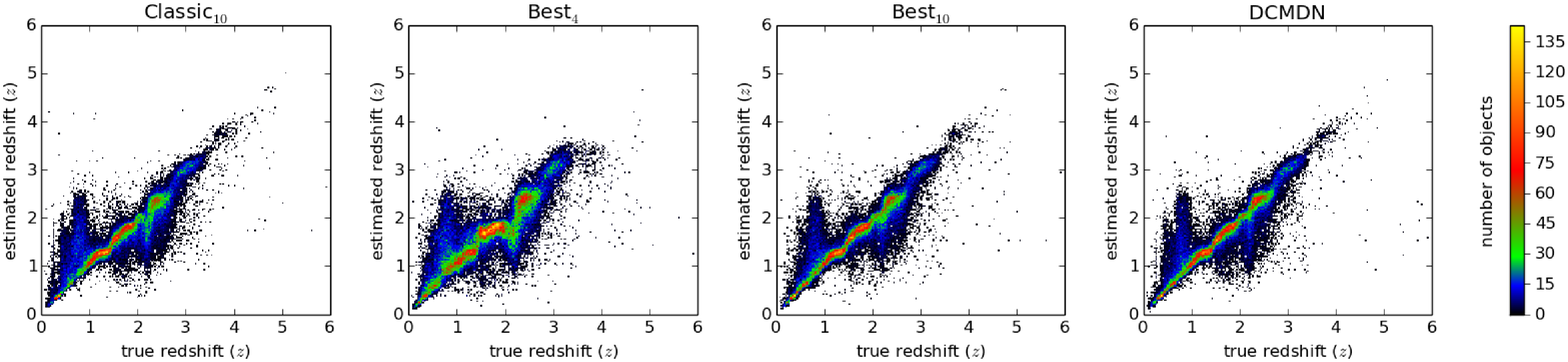}
\caption{Comparison of the spectroscopic (true) redshifts ($z_{spec}$) against the photometrically estimated redshifts ($z_{phot}$) of the different feature sets in experiment \textbf{DR7+9}.}\label{Fig:plots}
\end{figure*}

\begin{figure*}[ht]
\centering
\includegraphics[width=1.0\textwidth]{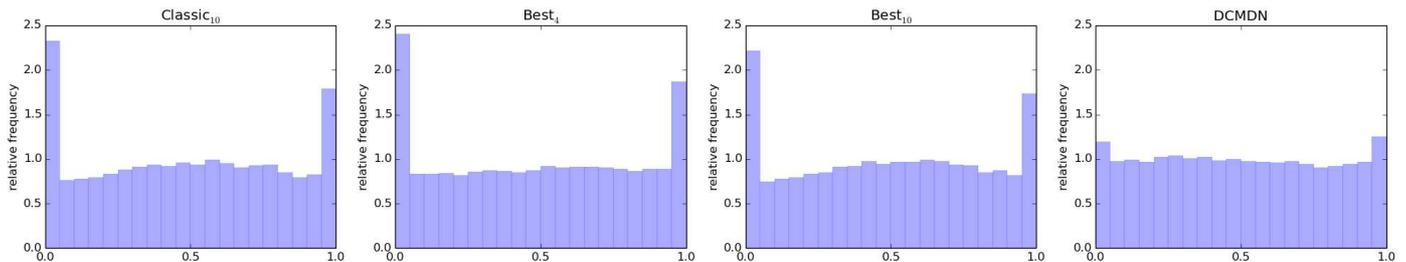}
\caption{PIT histograms for experiment \textbf{DR7+9} for the different features sets, as shown in Table~\ref{tab:crps_results}. Except the PIT of the DCMDN, all other feature sets generate results with significant outliers at the extrema.}\label{Fig:pits}
\end{figure*}

\begin{figure}
   \centering
   \includegraphics[width=1.0 \columnwidth]{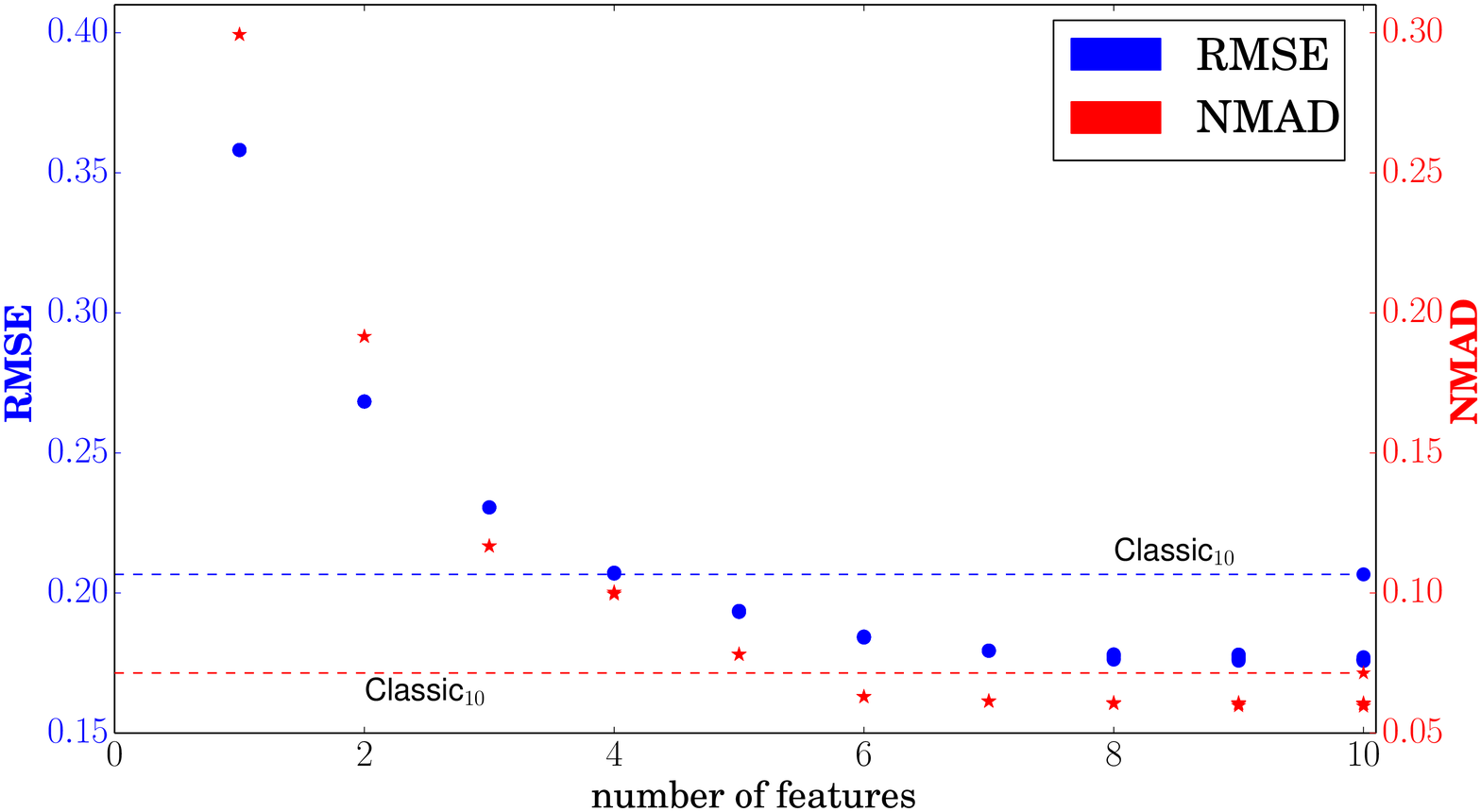}
   \caption{Comparison of model performance with regard to the number of used features. The root mean square error and normalized median absolute deviation of the results from the DR7+9 RF experiments are presented. As reference line the performance achieved with the  {\sffamily{Classic$_{10}$}} features is shown. As it can be seen, from the fourth feature on, the performance of the subsets outperforms the {\sffamily{Classic$_{10}$}} features. After the ninth feature, the improvement settles. When adding many more features, the performance will start to degrade.}\label{Fig:rmsTrend}
\end{figure}

\noindent
The best performing subset is shown in the fourth column of Table~\ref{tab:best_features} (DR7+9 subset), while in the third section of Table~\ref{tab:results_summary} results obtained with the RF experiments are given.
Moreover, in the right section of Table~\ref{tab:crps_results} the results with the CRPS as indicator are provided.
For this experiment we also report the $z_{spec}$ versus $z_{phot}$ plots in Fig.~\ref{Fig:plots}.
This classical representation visualizes the better concentration along the ideal diagonal for both the {\sffamily{Best$_{10}$}} features as well as the features derived through the DCMDN.
When using the features in a probabilistic context, the better performance with respect to outliers of the DCMDN can be observed (Fig.~\ref{Fig:pits}).
The probability integral transform \citep[PIT][]{2005MWRv..133.1098G} histograms show very similar performances for all the feature sets that were selected.
Besides the outliers, the estimates are sharp and well calibrated, exhibiting no difference in comparison to the results generated with the {\sffamily{Classic$_{10}$}} features.
This is a good indication that no systematic biases were added through the selection process.

Finally, the performance obtained with the {\sffamily{Classic$_{10}$}} features is compared to the ones achieved with the {\sffamily{Best$_{10}$}} features in a cumulative way.
In Fig.~\ref{Fig:rmsTrend}, the RMSE and the NMAD are plotted with respect to the number of features of the {\sffamily{Best$_{10}$}} set that were used.
This is important in order to show that starting with the $4th$ feature, the model reaches already a performance comparable with the {\sffamily{Classic$_{10}$}} features.
Originating in the random data sampling during the selection process, the resulting different feature subsets do not show obvious differences in the quality of the final performance.
In fact, the results obtained with the {\sffamily{Best$_{10}$}} subset are far better with respect to the performance obtained using the {\sffamily{Classic$_{10}$}} features and the DCMDN.
This is a confirmation of the quality and strength of the proposed method.

\section{Discussion}\label{discussion}

In the following subsections we discuss in detail the features found with the proposed method, the improvement in performance of the photometric redshift estimation models in comparison to the classic features, and the physical interpretation of the selected features.

\subsection{Features}

The results obtained from the feature selection process for the three experiments demonstrate that most of the information can be embedded in a limited number of features with respect to the initially generated amount of pairwise combinations.
The following four features have been selected and are in common between all the three experiments:
\vspace{2em}
\begin{itemize}
    \item{$r_{psf}-r_{petro}$}
    \item{$i_{psf}-i_{dev}$}
    \item{$\text{\ul{$i_{psf}$}}/\text{\ul{$z_{model}$}}$}
    \item{$\text{\ul{$r_{psf}$}}/\text{\ul{$i_{exp}$}}$.}
\end{itemize}

\noindent
This is a clear indicator that those features contain some essential information.
Besides noting that they encode \changeAdd{spatial and morphological characteristics}, we have no clear explanation.
Some features, as will be analysed in the next sections, can be clearly connected to physical processes occurring in the considered sources.
Other features are instead much harder to interpret, which demands a deeper analysis in the future.
Given that photometric redshifts are just used as a testbed for the proposed methodology, such an analysis is beyond the scope of this work.
A quick and shallow inspection of the features exhibits that the ratios and differences play a major role.
In Table~\ref{tab:treedefinition_dr79} for the experiment DR7+9 the different groups of features are highlighted using different background patterns.
This visually summarizes the dominant occurrence of those groups.
In fact, all the features except the $4th$ (errors) belong to one of these two groups.
Moreover, the individual branches of feature sets employ a feature of the same group for the first seven positions, showing a great stability in the composition of the branches.
The experiment based on the DR7+9 catalogue generates a much less complex structure of the tree of feature sets with respect to  experiments DR7a and DR7b.
Fewer branches and a reduced number of features are selected.
Reasons for this behaviour are the more complete redshift distribution of catalogue DR7+9 with respect to the other two and the improvement in SDSS photometry from DR7 to DR9.
This drives the model to find the required information in a reduced number of efficient features.
The analysis of the tree composition and features distribution can be done following the chord diagram shown in Fig.~\ref{Fig:chorddr79}.
The chord diagram is an optimal visualization tool for the description of a complex data structure.
In this diagram, every feature is associated to a specific colour, and starting from the first feature (A) it is possible to follow all the possible paths of the tree, depicting the different feature subsets.
Ordered from outside to inside, the external arcs represent the occurrences of a particular feature: the total percentage of the individual connections, the numbers and sources of connections entering, and the numbers and targets of connections exiting.
Therefore, the chord diagram, coupled with Table~\ref{tab:treedefinition_dr79}, gives a clear description of the structure and composition of the tree of features.
In addition, in Table~\ref{tab:treedefinition_dr79} the same colour code as in the chord diagram is adopted, to identify the features and their distribution.
The chord diagram clearly visualizes that the feature trees split at feature C and later rejoin at feature F.
In comparison to the chord diagram obtained for Experiment DR7+9, the two chord diagrams for experiments DR7a and DR7b (see Appendix~\ref{Sec:additions}) immediately visualize the higher complexity of those trees. 
From Fig.~\ref{Fig:chorddr79} and Table~\ref{tab:treedefinition_dr79} it appears that, apart from a few exceptions, the selected features follow a precise scheme.
No classic colour indexes or any of the {\sffamily{Classic$_{10}$}} features have been chosen, while only differences between different magnitudes of the same band or differences between different type of magnitudes play a certain role.
The ratios have been all selected in the extinction-corrected version, except for the subcategory of the photometric ratios.
This can be understood considering that the latter are ratios between magnitudes of the same filter where the contribution of the extinction correction tends to cancel out.

Another relevant aspect in experiment DR7+9 is that all the $15$ features in the tree are exclusively a composition of magnitudes and their errors.
Neither radii nor ellipticities have been chosen during the selection process.
As only quasars have been used in the experiments, this introduces a bias to the selection process in favour of magnitudes and against shape-based features.
This is a clear indication that just the magnitudes are required to describe the objects and explore the parameter space in the setting of photometric redshift estimation.
Although photometric ratios are shape-related parameters, they express the ratio between the centred and the extended part of a component that can be interpreted as flux of the hosting galaxy.
Therefore, here a bias introduced by using quasars for the experiments cannot be observed.

It is remarkable that photometric errors are selected as features, given that there is no obvious physical relation between the redshift of the considered objects and the measurement errors reported by the photometric pipeline of SDSS.
Therefore it is important to consider how errors are derived in the SDSS, based on flux measurements \citep{1999AJ....118.1406L}.
Magnitude errors quantify the discrepancy between the fitted photometric model (psf, model, petrosian, etc.) and the observed pixel-wise distribution of spatially correlated fluxes with respect to the applied noise model.
Therefore, it is evident that the errors on the single magnitudes appear to be larger for fainter objects, a physical property that is directly correlated to distance.
In addition, the deviation of spatial flux distributions from the applied spatial photometric models are good morphological indicators; for example, the shape and size of the hosting galaxy are correlated with redshift.
The workflow adopted is able to capture these dependencies, selecting a composition of errors as an important feature of the best set.

Even though $4,520$ features were synthetically created by combining base features, only $15$ were selected in experiment DR7+9 ($19$ and $26$ for experiments DR7a and DR7b, respectively).
Furthermore, some features encode the same type of information with just subtle differences in composition.
It is remarkable that every feature that is built on magnitudes incorporates a PSF magnitude.
Moreover, the model and exponential magnitude in the SDSS are related\footnote{\url{http://classic.sdss.org/dr7/algorithms/photometry.html\#mag\_model}}, with the model magnitude being just the better fitting model \changeAdd{when comparing} an exponential and a de Vaucouleurs profile.
In the first stages of the selection process, the proposed algorithm does not select differing branches but identifies essential features to produce good results when photometrically estimating redshifts.
These observations are also valid for the results found in experiments DR7a and DR7b.

\subsection{Comparison of performance}

Using the RF, the validation experiments were carried out on every feature set.
The second subset, indicated as {\sffamily{Best$_{10}$}}, gave a slightly better performance than the others.
Even though we would not consider this as a substantial effect, we decided to choose this as our reference set.
It can be noticed from Fig.~\ref{Fig:rmsTrend} that from the $4th$ feature on, every subset delivers a performance comparable to the performance of all ten features in the {\sffamily{Classic$_{10}$}} set, with respect to the RMSE.
Consistently, the use of more than four features outperforms the {\sffamily{Classic$_{10}$}}, independently of the subset used.
Adding more features improves further the performance and the trend becomes asymptotic around the $9th$ feature.
At a certain point, adding many more features results in a degradation of the redshift estimation performance.
After the $8th$ feature, the contribution is of a minor nature.
Just to have a fair comparison to the {\sffamily{Classic$_{10}$}} features, we decided to pick the same number of ten features, even though a smaller number is sufficient to outperform the {\sffamily{Classic$_{10}$}} features.
The performance improvement is evident seeing the results reported in Table~\ref{tab:results_summary} and Fig.~\ref{Fig:plots}.
It is important to note that the CRPS results (Table~\ref{tab:crps_results}) confirm the performance shown with respect to the other scores.
When predicting PDFs instead of point estimates, the PIT histograms (Fig.~\ref{Fig:pits}) indicate the DCMDN as the best calibrated model.
This result is reasonable because the DCMDN is the only model trained using the CRPS as loss function, which is focused on the PDFs calibration.
The kNN and the RF are instead based on the optimization of point estimates using the RMSE.
Therefore, the calibration of the PDFs estimated using the DCMDN is superior.
The use of such a probabilistic model is helpful to handle the presence of extreme outliers, since it is not based on the minimization of the RMSE, as discussed in \cite{2018A&A...609A.111D}.
The usage of PDFs allows us to identify objects with an ambiguous redshift distribution, while in a point estimation scenario, where just the mean of such a distribution would be considered, the estimates of those objects would result in extreme outliers.

Six features of the best subset are ratios of different magnitudes.
Three of them are plain ratios, while three are photometric ratios. 
Analysing the fourth column of Table~\ref{tab:best_features}, it appears that one of the components of these features is always a PSF magnitude, coupled with a model, petro, or exp magnitude.
Therefore, from the analysis of the results obtained, we can state that the reason for the performance improvement is not in the choice of some specific features, or in a particular subset of features, but in their type and in the combination of certain groups.

\begin{figure*}[!t]
   \centering
   \includegraphics[width=1.0\textwidth]{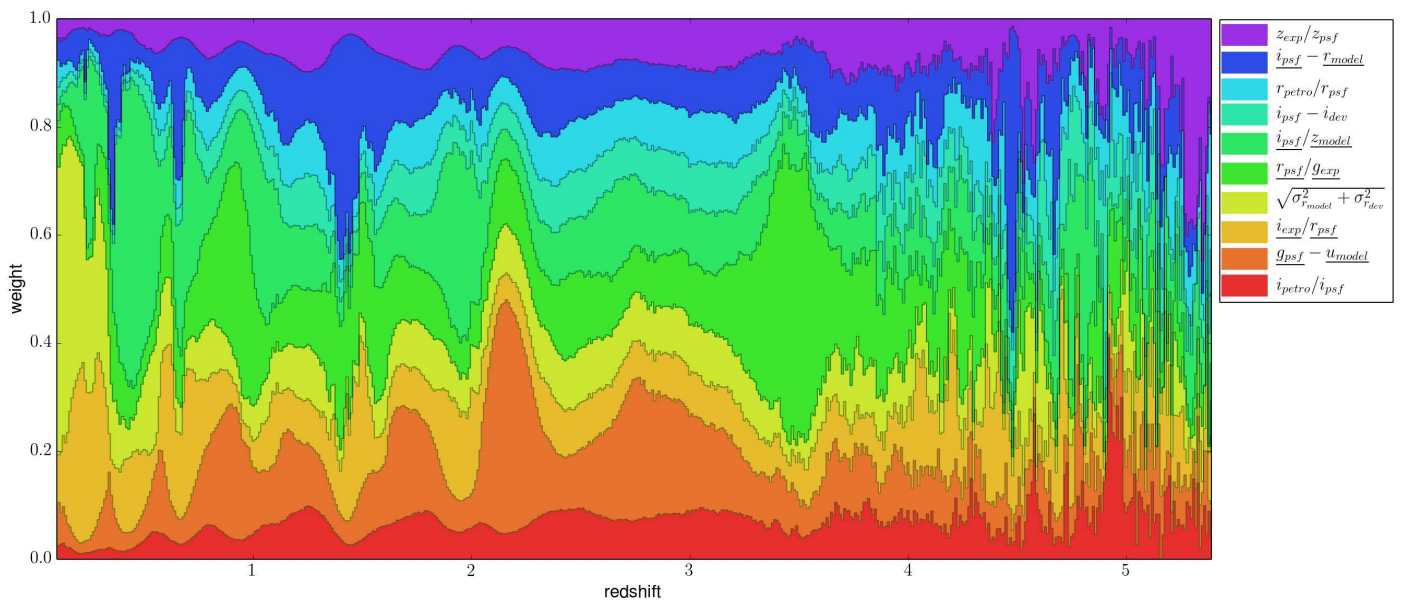}
   \caption{Importance of every feature of the {\sffamily{Best$_{10}$}} subset from experiment DR7+9. For a sliding redshift bin of $\Delta z=0.2$, the importance of every feature was calculated in a localized regression model based on the Gini index as utilized by the RF. The colour code used is the same adopted for the chord diagram in Fig.~\ref{Fig:chorddr79}.}\label{Fig:photoz_bins}
\end{figure*} 

All these aspects are clear indicators to demonstrate the following two conclusions.
The proposed method is highly stable, enabling us to derive subsets of features that are equivalently well-performing and similar, based on a common structure.
In this sense, the improvement with respect to the use of {\sffamily{Classic$_{10}$}} features is clear.
In order to prove the robustness of the proposed method, we performed some experiments using for each data-set the {\sffamily{Best$_{10}$}} features obtained with  the other two catalogues, as shown in Table~\ref{tab:cross_results}, and the results were almost as good as in the other cases.
The method captures the inherent structure of the physical properties of the sources, which is essential to provide good photometrically estimated redshifts for quasars.

\begin{table}\footnotesize
\centering
\begin{tabular}{|l|c|c|c|}
\hline
Exp. & catalogue DR7a & catalogue DR7b & catalogue DR7+9 \\
\hline 
DR7a & 0.124 & 0.146 & 0.176 \\
DR7b & 0.125 & 0.145 & 0.176 \\
DR7+9 & 0.124 & 0.147 & 0.174 \\
\hline\end{tabular}

\caption{Cross experiments performed with the RF, using the {\sffamily{Best$_{10}$}} sets obtained from every experiment with all the three catalogues. The results are expressed using the RMSE. It can be noticed the negligible difference of performance, for every catalogue, independently from the feature set used.} \label{tab:cross_results}
\end{table}

\begin{figure*}[p]
\centering
  \includegraphics[width=.98\textwidth]{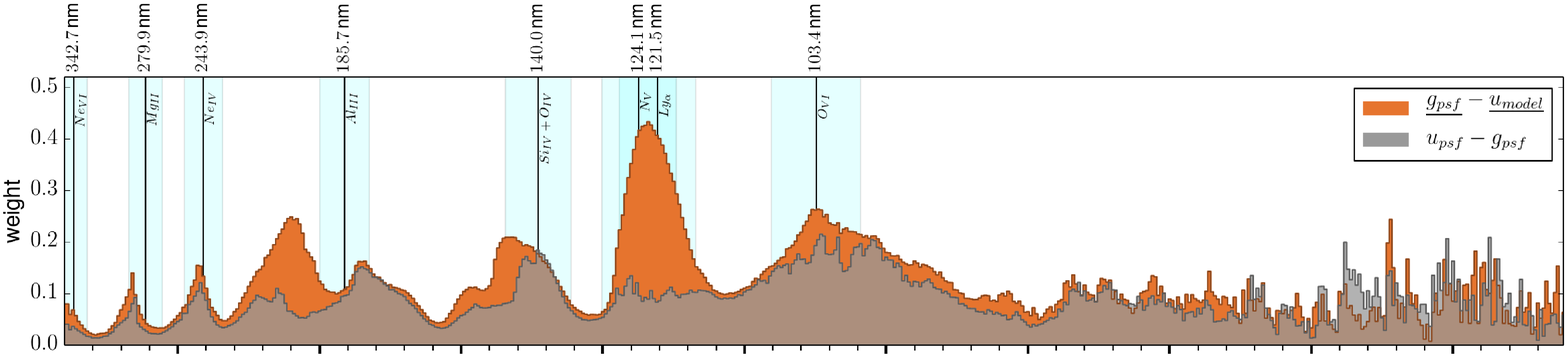}
  
  \vspace{0.5em}
  \includegraphics[width=.98\textwidth]{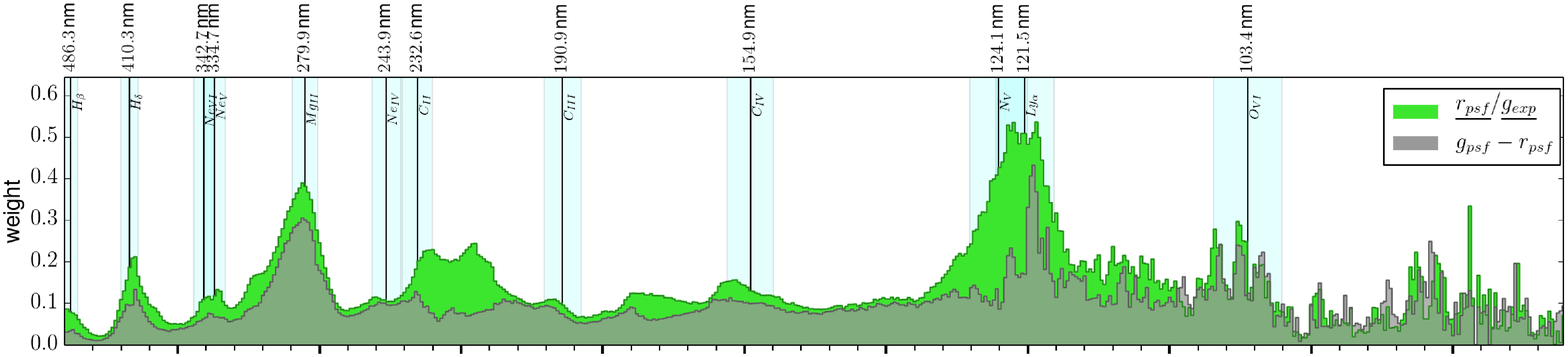}
  
  \vspace{0.5em}
  \includegraphics[width=.98\textwidth]{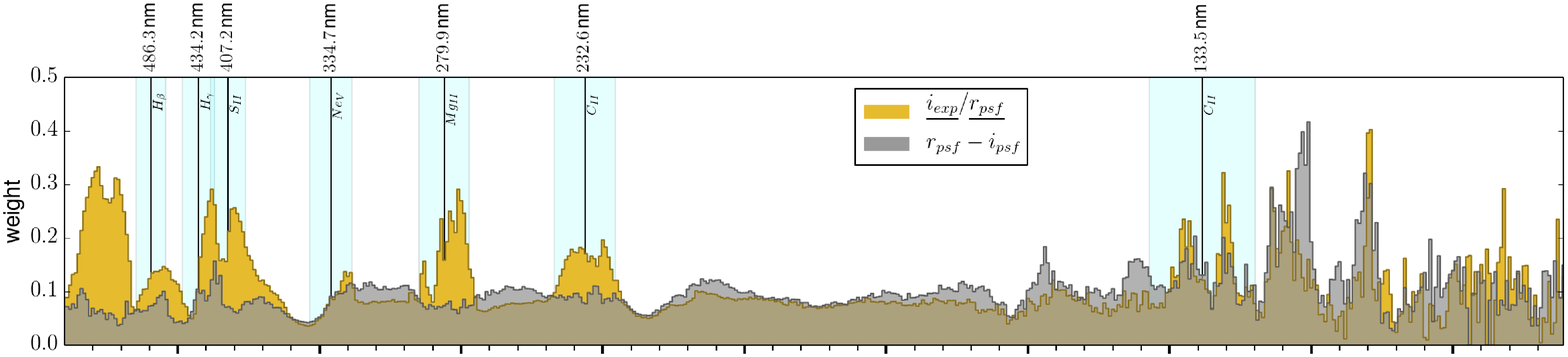}
  
  \vspace{0.5em}
  \includegraphics[width=0.98\textwidth]{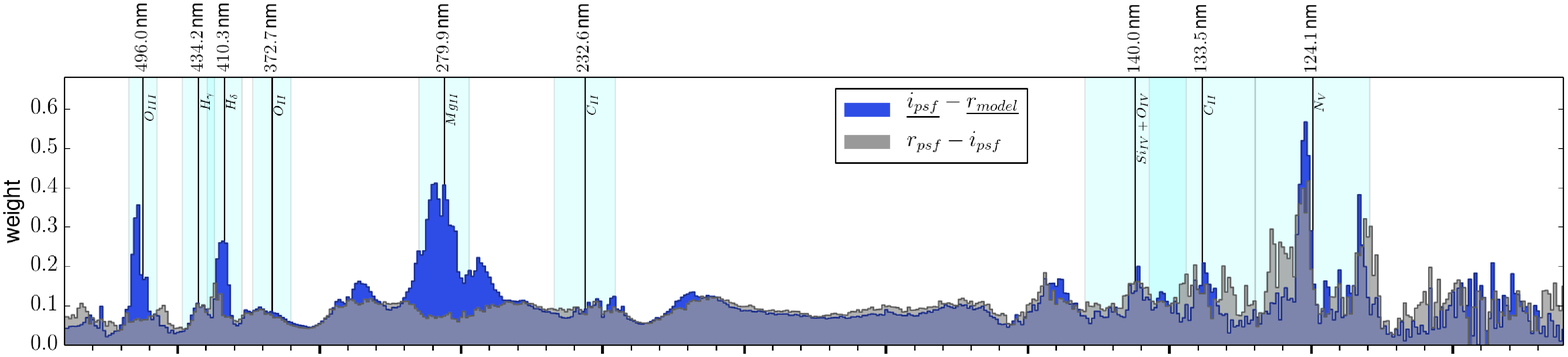}
  
  \vspace{0.5em}
  \includegraphics[width=0.98\textwidth]{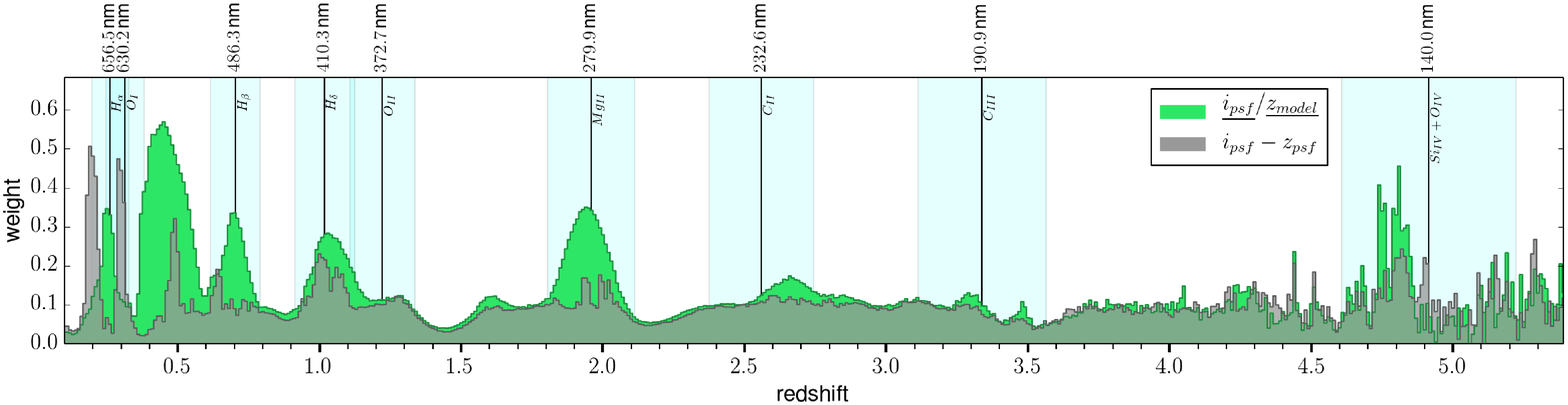}
\caption{Feature importance of the five features from the {\sffamily{Best$_{10}$}} set composed by magnitudes from neighbouring bands. As in Fig.~\ref{Fig:photoz_bins}, for a sliding redshift bin of $\Delta z=0.2$, the importance of every feature was calculated. The results are compared to the classic features using PSF magnitudes of the same bands. Based on the characteristics of the \emph{ugriz} filters,
the wavelengths indicating the start, centre, and end of the overlapping regions are used to overplot the positions of particular quasar emission lines using Eq.~\ref{Eq:zequation}.
The used colour code is the same as in Fig.~\ref{Fig:chorddr79}, while corresponding features of the {\sffamily{Classic$_{10}$}} set are always shown in grey.}
\label{Fig:importancevspsf}
\end{figure*}

\begin{figure*}[p]
\centering
  \includegraphics[width=0.98\textwidth]{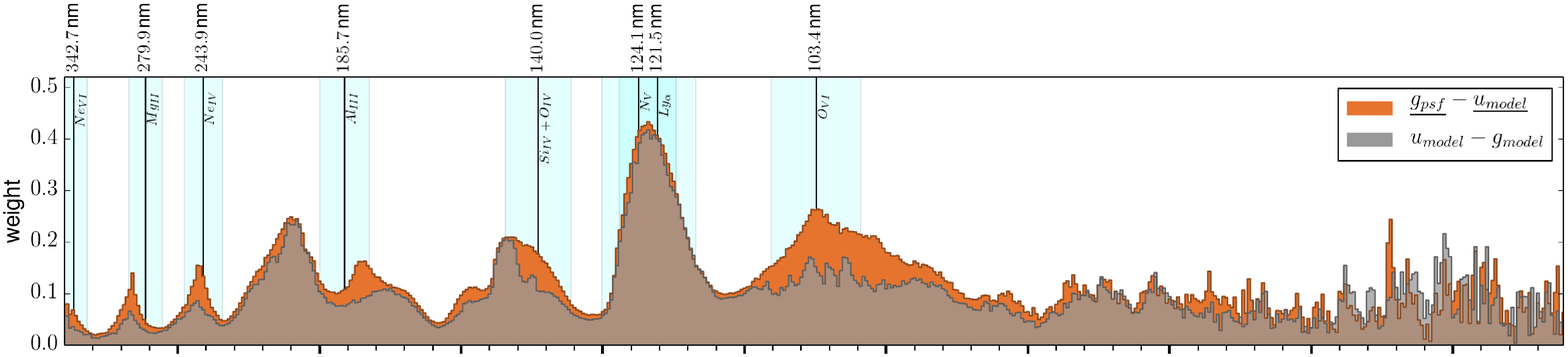}
  
  \vspace{0.5em}
  \includegraphics[width=0.98\textwidth]{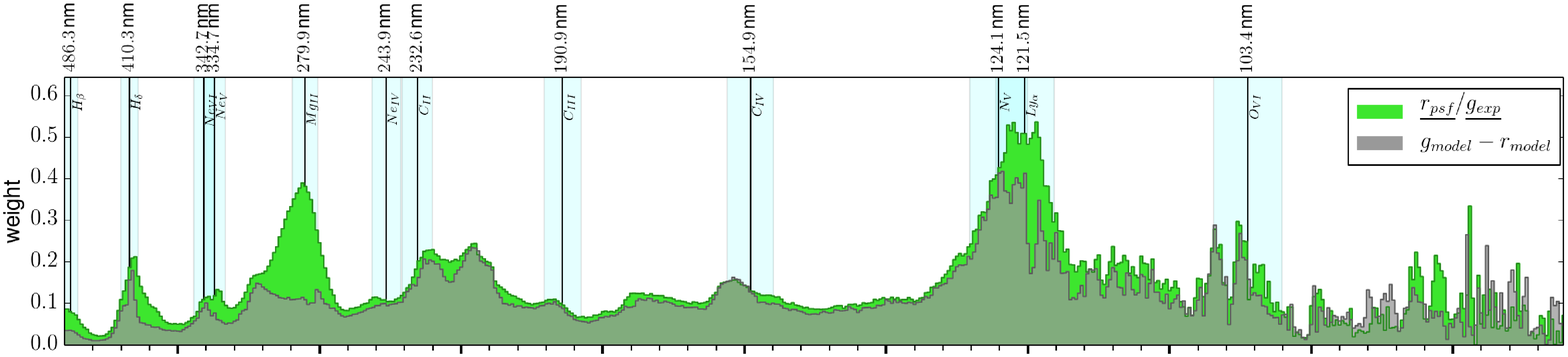}
  
  \vspace{0.5em}
  \includegraphics[width=.98\textwidth]{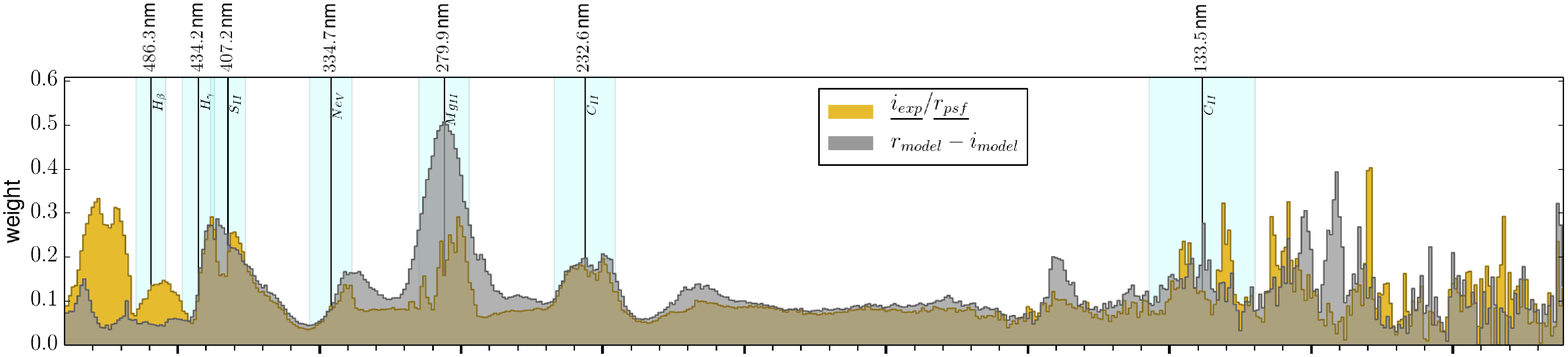}
  
  \vspace{0.5em}
  \includegraphics[width=.98\textwidth]{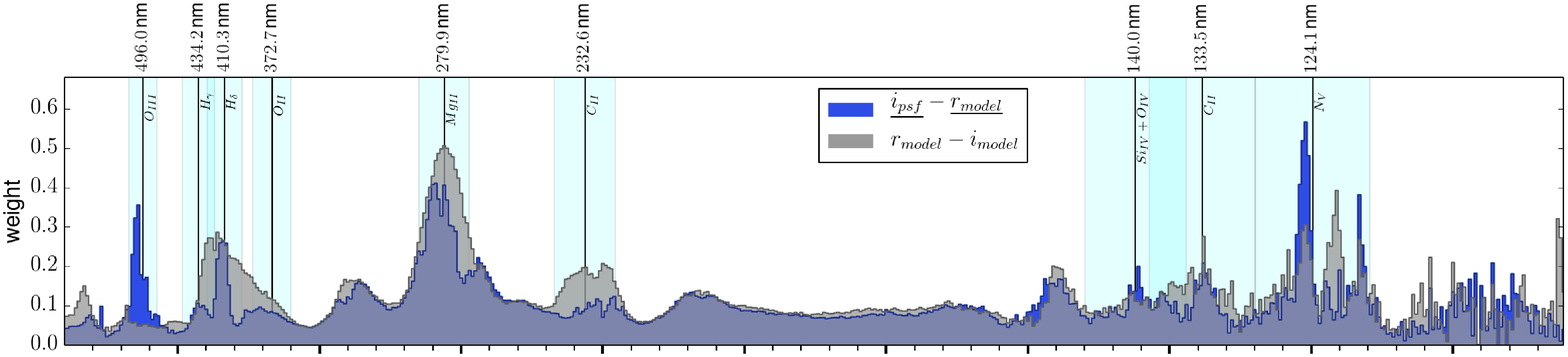}
  
  \vspace{0.5em}
  \includegraphics[width=.98\textwidth]{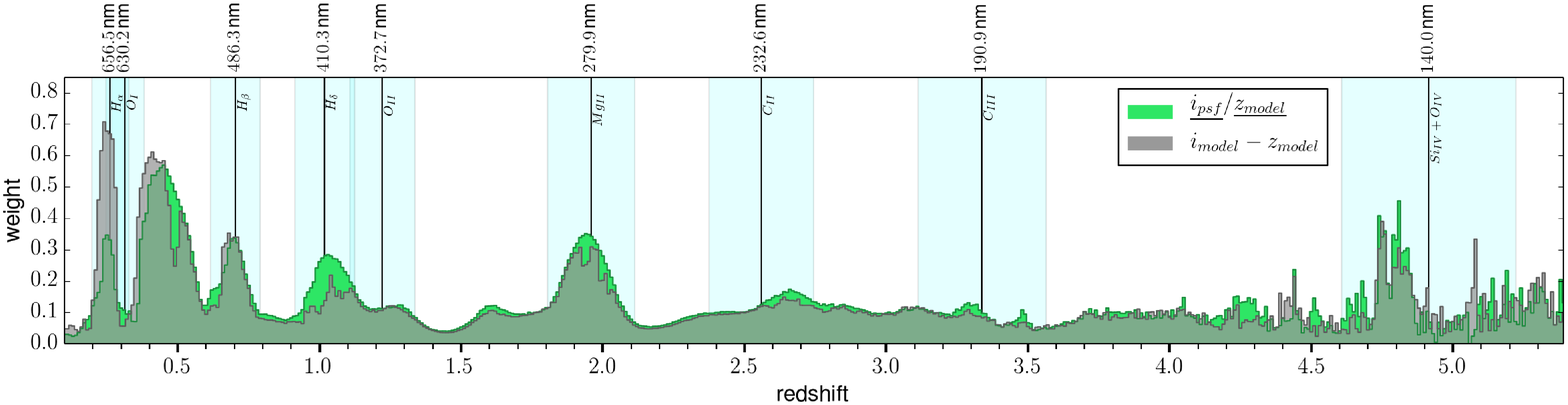}
\caption{Feature importance of the five features from the {\sffamily{Best$_{10}$}} set composed by magnitudes from neighbouring bands. As in Fig.~\ref{Fig:photoz_bins}, for a sliding redshift bin of $\Delta z=0.2$, the importance of every feature was calculated. The results are compared to the classic features using model magnitudes of the same bands. Based on the characteristics of the \emph{ugriz} filters,
the wavelengths indicating the start, centre, and end of the overlapping regions are used to overplot the positions of particular quasar emission lines using Eq.~\ref{Eq:zequation}.
The used colour code is the same as in Fig.~\ref{Fig:chorddr79}, while corresponding features of the {\sffamily{Classic$_{10}$}} set are always shown in grey.}
\label{Fig:importancevsmodel}
\end{figure*}

\subsection{Physical interpretation}

In contrast to deep learning models, feature-based approaches have the advantage of allowing an interpretation in a physical context.
Therefore the features selected by our approach are discussed in the following.
By analysing the importance of each feature of the {\sffamily{Best$_{10}$}} set in smaller redshift bins, the contribution of certain spectral features can be understood.
In Fig.~\ref{Fig:photoz_bins} the importance is presented for sliding bins of $\Delta z =0.2$ based on the Gini index (\citealt{cart84}).
The Gini index is used in the RF to perform the segmentation of the parameter space orthogonally to its dimensions at every node.
As all ten features contribute individually, the total contribution is normalized to one and the individual lines are presented in a cumulative way.
The relative importance of each feature clearly does not reflect their ordering, as they have been assembled by a forward feature selection algorithm.
In particular, the first feature of the best set does not show a dominant role when using multiple features.
When building a photometric regression model based on just a single feature, the concentration index in the \emph{i} band provides the best tracer for distance.
Therefore a concentration index in the \emph{i} band is consequently chosen in all the three experiments.
This selection is of course heavily biased by the distribution of our training objects with respect to redshift and by the fact that objects for training are selected based on the classification of the spectral template fitting of SDSS.
\definecolor{newgreen}{rgb}{0.0, 0.42, 0.24}
\begin{table}[b]
\centering\begin{tabular}{rl|l|l}
\multicolumn{2}{c|}{Position} & Feature & Score\\ 
\hline
$1$ &$ \begingroup    \color{newgreen}    \blacktriangle 1 \endgroup$& $\text{\ul{$g_{psf}$}}-\text{\ul{$u_{model}$}}$ & $0.424$ \\
$2$ &$ \begingroup    \color{red}    \blacktriangledown 1 \endgroup$& $i_{petro}/i_{psf}$ & $0.121$ \\
$3$ &$ \begingroup    \color{newgreen}    \blacktriangle 1 \endgroup$& $\sqrt{\sigma_{r_{model}}^2+\sigma_{r_{dev}}^2}$ & $0.092$ \\
$4$ &$ \begingroup    \color{red}    \blacktriangledown 1 \endgroup$& $\text{\ul{$i_{exp}$}}/\text{\ul{$r_{psf}$}}$ & $0.072$ \\
$5$ &$\bm{=\joinrel=}$& $\text{\ul{$r_{psf}$}}/\text{\ul{$g_{exp}$}}$ & $0.071$ \\
$6$ &$\bm{=\joinrel=}$& $\text{\ul{$i_{psf}$}}/\text{\ul{$z_{model}$}}$ & $0.064$ \\
$7$ &$ \begingroup    \color{newgreen}    \blacktriangle 2 \endgroup$& $\text{\ul{$i_{psf}$}}-\text{\ul{$r_{model}$}}$ & $0.062$ \\
$8$ &$ \begingroup    \color{red}    \blacktriangledown 1 \endgroup$& $i_{psf}-i_{dev}$ & $0.042$ \\
$9$ &$ \begingroup    \color{newgreen}    \blacktriangle 1 \endgroup$& $z_{exp}/z_{psf}$ & $0.026$ \\
$10$ &$ \begingroup    \color{red}    \blacktriangledown 2 \endgroup$& $r_{petro}/r_{psf}$ & $0.025$ \\
\hline
\end{tabular}
\caption{Features of the {\sffamily{Best$_{10}$}} set from experiment DR7+9, ordered by decreasing importance as expressed by the score of the RF based on the Gini criterion. The change with respect to the initially found ordering of the presented approach, and the RF score are reported, too.}\label{tab:feature_importance}
\end{table}
As soon as more photometric features are used, the spectral energy distribution and distinct spectral features are the dominant source of information for estimating the redshifts.
Those features are mainly ratios.
To use ratios instead of colours is a surprising fact, as in the literature colours are the usual choice for photometric redshift estimation models.
In Fig.~\ref{Fig:photoz_bins} one can inspect how the different features contribute at different redshift bins, building a well-performing model that covers the full redshift range.
Besides some very narrow redshift regions, no clear structure with preference of some photometric features can be observed at higher redshifts ($z > 4$).
This is due to the poor coverage of the training and validation data in that range.
The ordering of the features in the {\sffamily{Best$_{10}$}} set and their importance as shown in Fig.~\ref{Fig:photoz_bins} can be compared with the global feature importance as obtained from the RF experiment (Table~\ref{tab:feature_importance}).
The feature importance calculated on the overall redshift distribution gives different indications with respect to the bin-wise analysis, but it is quite consistent with the original order obtained from the feature selection.
This is a further demonstration of the stability and robustness of the proposed method.

The different behaviours and importance found for the features in the individual redshift bins can be partially explained by analysing distinct features in the spectral energy distribution.
By carefully inspecting the emission lines of quasars as reported by the SDSS spectral pipeline, a connection between some photometric features and emission lines could be found.
Those features that are composed of adjacent filter bands are very sensitive to spectral lines that are in the vicinity of the overlapping area of filter transmission curves.
This can be explained by a flipping of the feature, \changeAdd{for example positive or negative for colors and above or below one for ratios}.
Already a little shift of an emission line with respect to the redshift is enough to create a significant change in the feature space that is detected and utilized by the machine learning model.
Five features of the {\sffamily{Best$_{10}$}} share this characteristic.
Therefore the discussion with respect to emission lines is focused on selected features that are composed of magnitudes from neighbouring filter bands.
Using the well known relation
\begin{equation}\label{Eq:zequation}
    z = \frac{\lambda_{observed}}{\lambda_{emitted}} - 1 = \frac{\lambda_{filter\ intersection}}{\lambda_{qso\ emission\ line}} - 1,
\end{equation}
\noindent it is possible to calculate the redshift at which a specific emission line becomes traceable when using a certain filter combination.
The proposed features capture many distinct emission lines, showing peaks in the redshift bins where the lines appear.
This is shown in Figs.~\ref{Fig:importancevspsf} and \ref{Fig:importancevsmodel}, where the feature importance has been compared with the classic features of the corresponding bands.
To understand better the influence of the usage of magnitudes describing extended objects, both the PSF and the model magnitudes of the classic features where used for comparison.
In Fig.~\ref{Fig:importancevspsf} the comparison is performed with respect to PSF colours, while in Fig.~\ref{Fig:importancevsmodel} the same comparison is done with respect to model colours.
By using Eq.~\ref{Eq:zequation}, a selected set of spectral emission lines of quasars has been convolved with the corresponding filter characteristics to annotate the plots.
Besides the maximum of the overlapping region, the start and the end of the intersection are depicted.
We defined the upper and lower limits as the points at which the sensitivity of the filter curve is equal to $0.001$ in quantum efficiency.
It can be seen that many emission lines perfectly correspond to peaks in importance exhibited by the features of the {\sffamily{Best$_{10}$}} set.
This can be observed only partially for the classic features.

In particular, purely PSF or model magnitude-based colours have a different and often complementary contribution for several spectral lines.
This is due to the fact that either concentrated or extended characteristics of the analysed objects are considered.
The proposed features are more suitable than classic features to describe the peaks at distinct emission lines.
Considering the $N_{V}-Ly_{\alpha}$ lines for the {$\text{\ul{$g_{psf}$}}-\text{\ul{$u_{model}$}}$} feature, the comparison between the extended and concentrated classic features clearly indicates that an extended component of the source is captured via this feature.
Keeping in mind that a pixel size of $0.4"$ of the SDSS camera corresponds\footnote{using \citet{2006PASP..118.1711W} with $H_0=69.6, \Omega_M=0.286, \Omega_{DE}=0.714.$} at a redshift of $z \approx 2.2$ to $\approx 3.4\,$kpc, this is a clear indicator that the hosting galaxy is significantly contributing to the solution of the photometric redshift estimation model.
A similar behaviour can be observed for the $N_{V}-Ly_{\alpha}$ lines in the {$\text{\ul{$r_{psf}$}}/\text{\ul{$g_{exp}$}}$} feature, while the $Mg_{II}$ emission line mainly appears in the PSF colour.
Therefore the $Mg_{II}$ emission line can be considered to be more prominent in the central region of the objects.
Between the most notable lines, the Lyman-$\alpha$ and the Balmer series can be identified.
Other important lines found are the $C_{II}$, $C_{III}$, $C_{IV}$, $O_I, O_{II}, O_{III}, O_{VI}$ , and the $Mg_{II}$ lines.
Besides the identified peaks caused by specific emission lines, some peaks in weight stay unexplained.
Even though it is possible to distinguish between mostly spatially extended or concentrated characteristics of the objects, an association of a single emission line fails.
In those cases not the transition of a line between two filters but an overall shape relation is captured by the selected parameters.
As the selected features combine the strength of identifying line transitions as well as morphological characteristics, the resulting boost in performance of the photometric redshift estimation model can be well explained.
To explain the meaning of the selected features that use a combination of features extracted from the same photometric band and thereby describe a morphological structure of the source, further image-based investigations are necessary.
This proves that a model using the proposed feature selection approach is better able to exploit the information that represents the underlying \changeAdd{physical and morphological} structure as well as the processes going on in the sources.

\section{Conclusions}\label{conclusions}
In this work a method to select the best features for photometric redshift estimation is proposed.
The features are calculated via a greedy forward selection approach, in which the features are selected from a set of $4,520$ combinations based on the photometric and shape information stored in the SDSS DR7 and DR9 catalogues.
By randomly sampling the training data and running multiple kNN experiments, trees in which every branch constitutes a subset of features were generated for all the experiments.
The obtained branches were then validated using a RF model and compared to the results obtained using classic sets of features.
Moreover, the results were compared with a convolutional neural network based model, meant to automatically perform the feature extraction and selection.
Three experiments, based on different catalogues, were carried out.
The first catalogue was obtained selecting quasars from SDSS DR7 and applying photometric flags.
The second catalogue was composed of quasars from SDSS DR7 too, but without using photometric flags.
Finally, the third catalogue was made by mixing SDSS DR7 and DR9 quasars, in order to extend the redshift distribution.
We have shown that all the sets obtained in all the experiments outperform the {\sffamily{Classic$_{10}$}}, and in particular a best-performing branch has been identified for each catalogue.
The best sets also gave a better performance with respect to the automatic model \changeAdd{(even though the latter typically shows a better calibration and is less affected by outliers when predicting PDFs instead of point estimates)}.
The new best features obtained in the present work are not immediately comprehensible.
Further analysis shows a relation between the dominant features of the {\sffamily{Best$_{10}$}} set and the emission lines of quasars, which correspond to the peaks of importance of the different features along the redshift distribution.
The same analysis carried out on the {\sffamily{Classic$_{10}$}} features proves that the latter are not able to capture the same physical information as compactly as the selected features.
This explains why the results obtained with the proposed method are outstanding with respect to the ones obtained with the {\sffamily{Classic$_{10}$}} features.
Moreover, we demonstrate that the proposed features fill the redshift space in a complementary way, each adding information that is relevant in different redshift ranges.
The proposed method is highly stable, as shown from the distribution of the features and the groups to which they belong.
The experiments show that the useful information is concentrated in a reduced number of features, which are typically very different from the {\sffamily{Classic$_{10}$}}.
Furthermore, we verified that the difference in terms of performance with respect to the various sets is almost negligible.
This demonstrates that the true advantage with respect to the {\sffamily{Classic$_{10}$}} features is not given by the selected features themselves, but from their distribution and type in the specific set.
Therefore, the stability shown from the different branches, for example the common distribution scheme of the features, and the ability to better capture the underlying physical processes, explains the superior performance obtained.
The method is very general and could be applied to several tasks in astrophysics (and not only in astrophysics).
In the future we propose to apply it to different sources (i.e. galaxies with and without an active nuclei) in order to verify if the obtained features are general or if they are only related to the fine structure of the data itself and to this specific population of sources.
This includes the question of how much the processes of the active galactic nuclei dominate with respect to the processes in the surrounding galaxy the feature selection approach.
It goes without saying that this first step made in the interpretation of the new features could open new doors in the understanding of the physics of quasars with respect to \changeAdd{distance and age} by providing better and more precise tracers.
On the other hand, the method shows a different approach alternative to the application of deep learning, but also employing GPUs intensively.
Both approaches are meant to establish an affordable and well-performing method to precisely predict photometric redshifts, in light of the upcoming missions and instruments in the near future.

\begin{acknowledgements}
The authors gratefully acknowledge the support of the Klaus Tschira Foundation.
SC acknowledges support from the project ``Quasars at high redshift: physics and Cosmology'' financed by the ASI/INAF agreement 2017-14-H.0.
We would like to thank Nikos Gianniotis and Erica Hopkins for proofreading and commenting on this work.
Topcat has been used for this work \citep{2005ASPC..347...29T}.
The DCMDN model has been developed using Theano \citep{2016arXiv160502688T}.
To extract the image for the DCMDN we made use of HIPS \citep{2015A&A...578A.114F} and Montage \citep{2004ASPC..314..593B}.
Montage is funded by the National Science Foundation under Grant Number ACI-1440620, and was previously funded by the National Aeronautics and Space Administration's Earth Science Technology Office, Computation Technologies Project, under Cooperative Agreement Number NCC5-626 between NASA and the California Institute of Technology.
Funding for the SDSS and SDSS-II has been provided by the Alfred P. Sloan Foundation, the Participating Institutions, the National Science Foundation, the U.S. Department of Energy, the National Aeronautics and Space Administration, the Japanese Monbukagakusho, the Max Planck Society, and the Higher Education Funding Council for England. The SDSS Web Site is http://www.sdss.org/.
The SDSS is managed by the Astrophysical Research Consortium for the Participating Institutions. The Participating Institutions are the American Museum of Natural History, Astrophysical Institute Potsdam, University of Basel, University of Cambridge, Case Western Reserve University, University of Chicago, Drexel University, Fermilab, the Institute for Advanced Study, the Japan Participation Group, Johns Hopkins University, the Joint Institute for Nuclear Astrophysics, the Kavli Institute for Particle Astrophysics and Cosmology, the Korean Scientist Group, the Chinese Academy of Sciences (LAMOST), Los Alamos National Laboratory, the Max-Planck-Institute for Astronomy (MPIA), the Max-Planck-Institute for Astrophysics (MPA), New Mexico State University, Ohio State University, University of Pittsburgh, University of Portsmouth, Princeton University, the United States Naval Observatory, and the University of Washington.

\end{acknowledgements}

\bibliographystyle{aa}
\bibliography{references}

\appendix

\section{Additional tables and figures}\label{Sec:additions}
In this section, the additional tables for the features selection and the tree structure, together with the related chord diagrams for the experiments DR7a and DR7b are given.
A brief explanation of how to read a chord diagram follows.

\subsection{Chord diagram: how to read}
The chord diagram is a tool to visualize complex structures and relations in multidimensional data, which is arranged in a matrix shape.
The data are disposed in a circle and each element, in our case the features, is associated with a different colour.
The relations between the elements are expressed by ribbons which connect them, with a specific width related to the importance of that specific connection.
Therefore, the different ribbons can enter or exit from every arc, representing the features.
The chord diagrams utilized for this work are characterized by three external arcs for each feature.
Ordered from outside to inside, the external arcs represent the occurrences of a particular feature: the total percentage of the individual connections, the numbers and sources of connections entering, and the numbers and targets of connections exiting.
Therefore, starting from the first features indicated in the captions, it is possible to follow all the possible paths of the tree, depicting the different feature subsets and their global scheme.
Splitting points, joints, and complex interplay between feature groups can thereby be analyzed intuitively.

\section{Data}
The SDSS object IDs and coordinates of the extracted quasars for the three catalogues are available as supplementary information, as ASCII files.
The tables are only available in electronic form at the CDS via anonymous ftp to cdsarc.u-strasbg.fr (130.79.128.5) or via \url{http://cdsweb.u-strasbg.fr/cgi-bin/qcat?J/A+A/}.

\noindent \textbf{dr7a.csv} contains the SDSS object IDs and coordinates of the quasars for experiment DR7a.

\noindent \textbf{dr7b.csv} contains the SDSS object IDs and coordinates of the quasars for experiment DR7b.

\noindent \textbf{dr7+9.csv} contains the SDSS object IDs and coordinates of the quasars for experiment DR7+9.

\section{Code}
The code of the DCMDN model is available on the ASCL.\footnote{\url{http://www.ascl.net/ascl:1709.006}.}

\renewcommand{\arraystretch}{1.5}

\begin{table*}\resizebox{0.88\textwidth}{!}{\centering
\centering
\definecolor{colora}{RGB}{230,46,46}
\definecolor{colorb}{RGB}{230,92,46}
\definecolor{colorc}{RGB}{230,141,46}
\definecolor{colord}{RGB}{230,190,46}
\definecolor{colore}{RGB}{220,230,46}
\definecolor{colorf}{RGB}{174,230,46}
\definecolor{colorg}{RGB}{125,230,46}
\definecolor{colorh}{RGB}{77,230,46}
\definecolor{colori}{RGB}{46,230,64}
\definecolor{colorj}{RGB}{46,230,113}
\definecolor{colork}{RGB}{46,230,159}
\definecolor{colorl}{RGB}{46,230,208}
\definecolor{colorm}{RGB}{46,202,230}
\definecolor{colorn}{RGB}{46,153,230}
\definecolor{coloro}{RGB}{46,104,230}
\definecolor{colorp}{RGB}{46,58,230}
\definecolor{colorq}{RGB}{83,46,230}
\definecolor{colorr}{RGB}{132,46,230}
\definecolor{colors}{RGB}{181,46,230}
\definecolor{colort}{RGB}{230,46,230}

\begin{tikzpicture}
\matrix (m)[matrix of nodes, style={nodes={rectangle,minimum width=8em,  row sep=-\pgflinewidth, column sep=-\pgflinewidth}}, ampersand replacement =\&]
{
  {\bf id} \&
  {\bf Feature 1} \& {\bf Feature 2} \& {\bf  Feature 3} \&  {\bf Feature 4} \&  {\bf Feature 5} \&  {\bf Feature 6} \&  {\bf Feature 7} \&  {\bf Feature 8} \& {\bf  Feature 9} \& {\bf  Feature 10} \\
\hline
$1  $\&\raisebox{1.2em}{$$}\&\raisebox{1.2em}{$$}\&\raisebox{1.2em}{$$}\&\raisebox{1.2em}{$$}\&\raisebox{1.2em}{$$}\&\raisebox{1.2em}{$$}\&\raisebox{1.2em}{$$}\&\raisebox{1.2em}{$$}\raisebox{-1.2em}{$$}\colorbox{white}{$  { \sigma_{g_{model}}}  $}\&\raisebox{1.2em}{$$}\raisebox{-1.2em}{$$}\colorbox{white}{$   { z_{model}/z_{psf} }  $}\&\raisebox{1.2em}{$$}\colorbox{white}{$  { i_{psf}-i_{dev}}$}\\
$2  $\&$ $\&$ $\&$ $\&$ $\&\colorbox{white}{$  {\text{\ul{$g_{psf}$}}/\text{\ul{$i_{exp}$}} }  $}\&\colorbox{white}{$ { \text{\ul{$i_{psf}$}}/\text{\ul{$z_{model}$}} }  $}\&$ $\&\raisebox{1.2em}{$$}\raisebox{-1.2em}{$$}\colorbox{white}{$   { z_{model}/z_{psf} }  $}\&\raisebox{1.2em}{$$}\raisebox{-1.2em}{$$}\colorbox{white}{$  { \sigma_{g_{model}}}  $}\&\raisebox{-1.2em}{$$}\\
$3  $\&$ $\&$ $\&$ $\&$ $\&$ $\&$ $\&$ $\&\raisebox{1.2em}{$$}\&\raisebox{1.2em}{$$}\&\raisebox{1.2em}{$$}\raisebox{-1.2em}{$$}\colorbox{white}{$  { \text{\ul{$g_{petro}$}}/\text{\ul{$r_{petro}$}}}$}\\
$4  $\&$ $\&$ $\&$ $\&\colorbox{white}{${ i_{psf}/i_{model} }  $}\&\raisebox{-1.2em}{$$}\&\raisebox{-1.2em}{$$}\&$ $\&\colorbox{white}{$  { \sigma_{g_{model}}}  $}\&\colorbox{white}{$   { z_{model}/z_{psf} }  $}\&\raisebox{1.2em}{$$}\raisebox{-1.2em}{$$}\colorbox{white}{$  { i_{dev}/i_{psf}}$}\\
$5  $\&$ $\&$ $\&$ $\&$ $\&\raisebox{1.2em}{$$}\&\raisebox{1.2em}{$$}\&\colorbox{white}{${ i_{psf}-i_{petro} }  $}\&\raisebox{-1.2em}{$$}\&\raisebox{-1.2em}{$$}\&\raisebox{1.2em}{$$}\\
$6  $\&$ $\&$ $\&$ $\&$ $\&\colorbox{white}{$  { \text{\ul{$i_{psf}$}}/\text{\ul{$z_{model}$}} }  $}\&\colorbox{white}{$  { \text{\ul{$g_{petro}$}}/\text{\ul{$r_{petro}$}} }  $}\&$ $\&\raisebox{1.2em}{$$}\&\raisebox{1.2em}{$$}\raisebox{-1.2em}{$$}\colorbox{white}{$  { \sigma_{g_{model}}}  $}\&\colorbox{white}{$  { i_{psf}-i_{dev}}$}\\
$7  $\&$ $\&$ $\&$ $\&$ $\&$ $\&$ $\&$ $\&\colorbox{white}{$   { z_{model}/z_{psf} }  $}\&\raisebox{1.2em}{$$}\colorbox{white}{$  { \sqrt{\sigma_{r_{model}}^2 + \sigma_{g_{dev}}^2} }  $}\&\raisebox{-1.2em}{$$}\\
$8  $\&   \colorbox{white}{${ i_{psf}/i_{exp} }  $}  \&$ $\&\colorbox{white}{${ \text{\ul{$r_{psf}$}}/\text{\ul{$i_{exp}$}} }  $}\&\raisebox{-1.2em}{$$}\&\raisebox{-1.2em}{$$}\&\raisebox{-1.2em}{$$}\&$ $\&\raisebox{-1.2em}{$$}\&\raisebox{-1.2em}{$$}\&\raisebox{1.2em}{$$}\raisebox{-1.2em}{$$}\colorbox{white}{$  { i_{dev}/i_{psf}}$}\\
$9  $\&$ $\&$ $\&$ $\&\raisebox{1.2em}{$$}\&\raisebox{1.2em}{$$}\&\raisebox{1.2em}{$$}\&\raisebox{-1.2em}{$$}\&\raisebox{1.2em}{$$}\raisebox{-1.2em}{$$}\colorbox{white}{$   { i_{dev}/i_{psf} }  $}\&\raisebox{1.2em}{$$}\raisebox{-1.2em}{$$}\colorbox{white}{$  { \sigma_{g_{model}}}  $}\&\raisebox{1.2em}{$$}\\
$10  $\&$ $\&$ $\&$ $\&$ $\&\colorbox{white}{$  { \text{\ul{$g_{psf}$}}/\text{\ul{$i_{exp}$}} }  $}\&\colorbox{white}{$  { \text{\ul{$i_{psf}$}}/\text{\ul{$z_{model}$}} }  $}\&\raisebox{1.2em}{$$}\colorbox{white}{$  { i_{dev}/i_{psf} }  $}\&\raisebox{1.2em}{$$}\raisebox{-1.2em}{$$}\colorbox{white}{$  { \sigma_{g_{model}}}  $}\&\raisebox{1.2em}{$$}\colorbox{white}{$  { r_{psf}-r_{petro} }  $}\&\colorbox{white}{$  { \text{\ul{$g_{petro}$}}/\text{\ul{$r_{petro}$}}}$}\\
$11  $\&$ $\&\colorbox{white}{$  {\text{\ul{$g_{psf}$}}/\text{\ul{$u_{model}$}} }  $}\&$ $\&$ $\&\raisebox{-1.2em}{$$}\&\raisebox{-1.2em}{$$}\&\raisebox{-1.2em}{$$}\&\raisebox{1.2em}{$$}\raisebox{-1.2em}{$$}\colorbox{white}{$  { \sqrt{\sigma_{r_{model}}^2 + \sigma_{g_{dev}}^2} }  $}\&\raisebox{-1.2em}{$$}\&\raisebox{-1.2em}{$$}\\
$12  $\&$ $\&$ $\&$ $\&\colorbox{white}{$ {z_{model}/z_{psf}}  $}\&\raisebox{1.2em}{$$}\&\raisebox{1.2em}{$$}\&\raisebox{1.2em}{$$}\&\raisebox{1.2em}{$$}\raisebox{-1.2em}{$$}\colorbox{white}{$  { \sigma_{g_{model}}}  $}\&\raisebox{1.2em}{$$}\raisebox{-1.2em}{$$}\colorbox{white}{$   { z_{model}/z_{psf} }  $}\&\raisebox{1.2em}{$$}\raisebox{-1.2em}{$$}\colorbox{white}{$  { i_{psf}-i_{dev}}$}\\
$13$\&$ $\&$ $\&$ $\&$ $\&$ $\&$ $\&$ $\&\raisebox{1.2em}{$$}\&\raisebox{1.2em}{$$}\colorbox{white}{$  { \sigma_{g_{model}}}  $}\&\raisebox{1.2em}{$$}\raisebox{-1.2em}{$$}\colorbox{white}{$  { r_{psf}-r_{petro}}$}\\
$14$\&$ $\&$ $\&$ $\&$ $\&$ $\&$ $\&\colorbox{white}{$ { i_{psf}-i_{petro} }  $}\&\colorbox{white}{$   { \text{\ul{$i_{dev}$}}/\text{\ul{$i_{psf}$}} }  $}\&\raisebox{-1.2em}{$$}\&\raisebox{1.2em}{$$}\raisebox{-1.4em}{$$}\colorbox{white}{$   { r_{dev}/r_{psf}}$}\\
$15$\&$ $\&$ $\&$ $\&$ $\&\colorbox{white}{$  { \text{\ul{$i_{psf}$}}/\text{\ul{$z_{model}$}} }  $}\&\colorbox{white}{$  { \text{\ul{$g_{petro}$}}/\text{\ul{$r_{petro}$}} }  $}\&$ $\&$ $\&\raisebox{1.2em}{$$}\colorbox{white}{$  { \sqrt{\sigma_{r_{model}}^2 + \sigma_{g_{dev}}^2} }  $}\&\raisebox{1.4em}{$$}\raisebox{-1.2em}{$$}\colorbox{white}{$  { r_{psf}-r_{petro}}$}\\
$16  $\&\raisebox{-1.2em}{$$}\&$ $\&\raisebox{-1.2em}{$$}\&\raisebox{-1.2em}{$$}\&$ $\&$ $\&\raisebox{-1.2em}{$$}\&\raisebox{-1.2em}{$$}\&\raisebox{-1.2em}{$$}\&\raisebox{1.2em}{$$}\raisebox{-1.2em}{$$}\colorbox{white}{$   { r_{dev}/r_{psf}}$}\\
$17  $\&\raisebox{1.2em}{$$}\&$ $\&\raisebox{1.2em}{$$}\&\raisebox{1.2em}{$$}\&$ $\&$ $\&\raisebox{1.2em}{$$}\&\raisebox{1.2em}{$$}\raisebox{-1.2em}{$$}\colorbox{white}{$  { \sigma_{g_{model}}}  $}\&\raisebox{1.2em}{$$}\&\raisebox{1.2em}{$$}\\
$18  $\&$ $\&$ $\&$ $\&$ $\&\raisebox{-1.2em}{$$}\&\raisebox{-1.2em}{$$}\&$ $\&\raisebox{1.2em}{$$}\colorbox{white}{$  { \sqrt{\sigma_{r_{model}}^2 + \sigma_{g_{dev}}^2} }  $}\&\colorbox{white}{$   { z_{model}/z_{psf} }  $}\&\colorbox{white}{$  { i_{psf}-i_{petro}}$}\\
$19  $\&\colorbox{white}{$ { i_{psf}/i_{model} }  $}\&$ $\&\colorbox{white}{$ { \text{\ul{$r_{psf}$}}/\text{\ul{$i_{model}$}} }  $}\&\colorbox{white}{$ { \text{\ul{$i_{dev}$}}/\text{\ul{$i_{psf}$}} }  $}\&\raisebox{1.2em}{$$}\colorbox{white}{$  { \text{\ul{$r_{psf}$}}/\text{\ul{$g_{model}$}} }  $}\&\raisebox{1.2em}{$$}\&\colorbox{white}{$ {r_{psf}-r_{petro}}  $}\&\raisebox{-1.2em}{$$}\&$ $\&$  $\\
$  20*   $\&$ $\&$ $\&$ $\&$ $\&\raisebox{-1.2em}{$$}\&\colorbox{white}{$  { \text{\ul{$i_{psf}$}}/\text{\ul{$z_{model}$}} }  $}\&$ $\&\raisebox{1.2em}{$$}\raisebox{-1.2em}{$$}\colorbox{white}{$  { \sqrt{\sigma_{r_{model}}^2 + \sigma_{g_{exp}}^2} }  $}\&\raisebox{-1.2em}{$$}\&\raisebox{-1.2em}{$$}\\
$21  $\&$ $\&$ $\&$ $\&$ $\&\raisebox{1.2em}{$$}\colorbox{white}{$  { \text{\ul{$g_{psf}$}}/\text{\ul{$i_{exp}$}} }  $}\&$ $\&$ $\&\raisebox{1.2em}{$$}\colorbox{white}{$  { \sigma_{g_{model}}}  $}\&\raisebox{1.2em}{$$}\colorbox{white}{$  { \text{\ul{$g_{petro}$}}/\text{\ul{$r_{petro}$}} }  $}\&\raisebox{1.2em}{$$}\raisebox{-1.2em}{$$}\colorbox{white}{$   { z_{model}/z_{psf}}$}\\
$22  $\&\raisebox{-1.2em}{$$}\&\raisebox{-1.2em}{$$}\&\raisebox{-1.2em}{$$}\&\raisebox{-1.2em}{$$}\&\raisebox{-1.2em}{$$}\&\raisebox{-1.2em}{$$}\&\raisebox{-1.2em}{$$}\&\raisebox{-1.2em}{$$}\&\raisebox{-1.2em}{$$}\&\raisebox{1.2em}{$$}\raisebox{-1.2em}{$$}\colorbox{white}{ ${ i_{psf}-i_{petro}}$}\\
};

\begin{pgfonlayer}{myback}
\fillpattern[dots][colorh]{m-2-2}{m-17-2}
\fillpattern[dots][colorj]{m-18-2}{m-23-2}
\fillpattern[vertical lines][colori]{m-2-3}{m-23-3}
\fillpattern[vertical lines][colorg]{m-2-4}{m-17-4}
\fillpattern[vertical lines][colorl]{m-18-4}{m-23-4}
\fillpattern[dots][colorj]{m-2-5}{m-9-5}
\fillpattern[dots][colorf]{m-10-5}{m-17-5}
\fillpattern[dots][colore]{m-18-5}{m-23-5}
\fillpattern[vertical lines][colork]{m-2-6}{m-5-6}
\fillpattern[vertical lines][coloro]{m-6-6}{m-9-6}
\fillpattern[vertical lines][colork]{m-10-6}{m-12-6}
\fillpattern[vertical lines][coloro]{m-13-6}{m-19-6}
\fillpattern[vertical lines][colorn]{m-20-6}{m-21-6}
\fillpattern[vertical lines][colork]{m-22-6}{m-23-6}
\fillpattern[vertical lines][coloro]{m-2-7}{m-5-7}
\fillpattern[vertical lines][colorp]{m-6-7}{m-9-7}
\fillpattern[vertical lines][coloro]{m-10-7}{m-12-7}
\fillpattern[vertical lines][colorp]{m-13-7}{m-19-7}
\fillpattern[vertical lines][coloro]{m-20-7}{m-23-7}
\fillpattern[horizontal lines][colora]{m-2-8}{m-10-8}
\fillpattern[dots][colore]{m-11-8}{m-12-8}
\fillpattern[horizontal lines][colora]{m-13-8}{m-17-8}
\fillpattern[horizontal lines][colorq]{m-18-8}{m-23-8}
\fillpattern[north west lines][colorc]{m-2-9}{m-2-9}
\fillpattern[dots][colorf]{m-3-9}{m-3-9}
\fillpattern[north west lines][colorc]{m-4-9}{m-6-9}
\fillpattern[dots][colorf]{m-7-9}{m-9-9}
\fillpattern[dots][colore]{m-10-9}{m-10-9}
\fillpattern[north west lines][colorc]{m-11-9}{m-11-9}
\fillpattern[north west lines][colorr]{m-12-9}{m-12-9}
\fillpattern[north west lines][colorc]{m-13-9}{m-13-9}
\fillpattern[dots][colore]{m-14-9}{m-17-9}
\fillpattern[north west lines][colorc]{m-18-9}{m-18-9}
\fillpattern[north west lines][colorr]{m-19-9}{m-20-9}
\fillpattern[north west lines][colort]{m-21-9}{m-21-9}
\fillpattern[north west lines][colorc]{m-22-9}{m-23-9}
\fillpattern[dots][colorf]{m-2-10}{m-2-10}
\fillpattern[north west lines][colorc]{m-3-10}{m-3-10}
\fillpattern[dots][colorf]{m-4-10}{m-6-10}
\fillpattern[north west lines][colorc]{m-7-10}{m-7-10}
\fillpattern[north west lines][colorr]{m-8-10}{m-9-10}
\fillpattern[north west lines][colorc]{m-10-10}{m-10-10}
\fillpattern[horizontal lines][colorq]{m-11-10}{m-12-10}
\fillpattern[dots][colorf]{m-13-10}{m-13-10}
\fillpattern[north west lines][colorc]{m-14-10}{m-15-10}
\fillpattern[north west lines][colorr]{m-16-10}{m-17-10}
\fillpattern[dots][colorf]{m-18-10}{m-21-10}
\fillpattern[vertical lines][colorp]{m-22-10}{m-23-10}
\fillpattern[horizontal lines][colord]{m-2-11}{m-3-11}
\fillpattern[vertical lines][colorp]{m-4-11}{m-4-11}
\fillpattern[horizontal lines][colore]{m-5-11}{m-5-11}
\fillpattern[horizontal lines][colord]{m-6-11}{m-8-11}
\fillpattern[horizontal lines][colore]{m-9-11}{m-9-11}
\fillpattern[vertical lines][colorp]{m-10-11}{m-12-11}
\fillpattern[horizontal lines][colord]{m-13-11}{m-13-11}
\fillpattern[horizontal lines][colorq]{m-14-11}{m-14-11}
\fillpattern[dots][colors]{m-15-11}{m-15-11}
\fillpattern[horizontal lines][colorq]{m-16-11}{m-16-11}
\fillpattern[dots][colors]{m-17-11}{m-17-11}
\fillpattern[horizontal lines][colora]{m-18-11}{m-21-11}
\fillpattern[dots][colorf]{m-22-11}{m-22-11}
\fillpattern[horizontal lines][colora]{m-23-11}{m-23-11}
\end{pgfonlayer}
\end{tikzpicture}
}
\caption{Detailed feature branches obtained from the feature selection for the experiment DR7a. The $20th$ branch, indicated with the $*$ symbol, is the best performing subset with respect to the experiments using the RF. The \emph{ratios} and \emph{photometric ratios} are indicated, respectively, with vertical lines and dots. The \emph{differences} are marked with horizontal lines and the \emph{errors} with north west lines. The color code for the features is the same as shown in the chord diagram in Fig.~\ref{Fig:chorddr7a}.}\label{tab:treedefinition_dr7}\end{table*}

\begin{figure*}
\definecolor{colora}{RGB}{230,46,46}
\definecolor{colorb}{RGB}{230,92,46}
\definecolor{colorc}{RGB}{230,141,46}
\definecolor{colord}{RGB}{230,190,46}
\definecolor{colore}{RGB}{220,230,46}
\definecolor{colorf}{RGB}{174,230,46}
\definecolor{colorg}{RGB}{125,230,46}
\definecolor{colorh}{RGB}{77,230,46}
\definecolor{colori}{RGB}{46,230,64}
\definecolor{colorj}{RGB}{46,230,113}
\definecolor{colork}{RGB}{46,230,159}
\definecolor{colorl}{RGB}{46,230,208}
\definecolor{colorm}{RGB}{46,202,230}
\definecolor{colorn}{RGB}{46,153,230}
\definecolor{coloro}{RGB}{46,104,230}
\definecolor{colorp}{RGB}{46,58,230}
\definecolor{colorq}{RGB}{83,46,230}
\definecolor{colorr}{RGB}{132,46,230}
\definecolor{colors}{RGB}{181,46,230}
\definecolor{colort}{RGB}{230,46,230}
\center{\textbf{Experiment DR7a}}\\
\begin{minipage}[]{0.60\textwidth}
\centering
\includegraphics[width=0.99\textwidth]{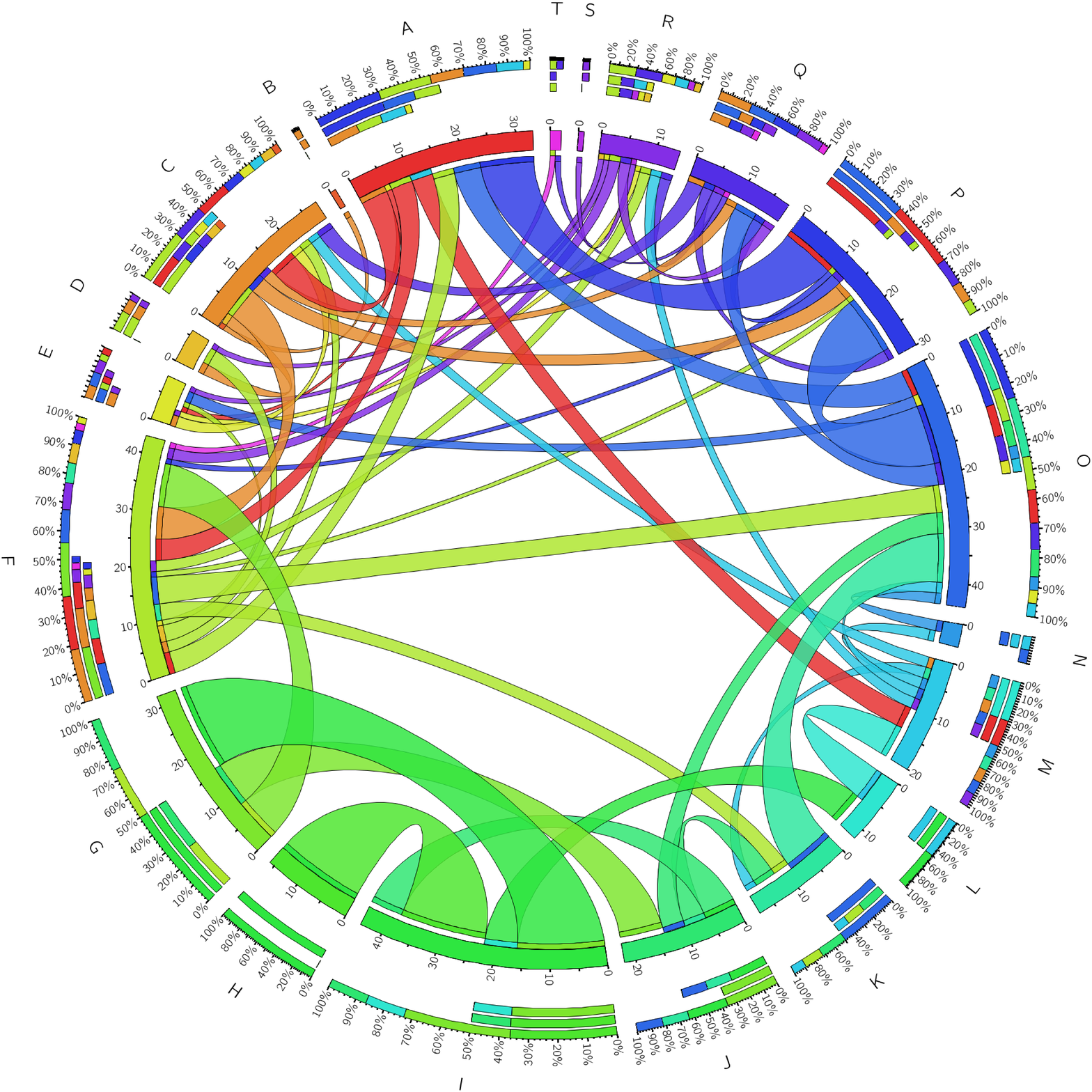}
\end{minipage}\hfill
\begin{minipage}[]{0.395\textwidth}
\renewcommand{\arraystretch}{1.5}
\tiny
\begin{tabular}{ccc}
 & name & feature\\
\colsquare{colora}&{\sffamily A}        &       $i_{psf}-i_{petro}$     \\
\colsquare{colorb}&{\sffamily B}        &       $r_{dev}/r_{psf}$       \\
\colsquare{colorc}&{\sffamily C}        &       $\sigma_{g_{model}}$    \\
\colsquare{colord}&{\sffamily D}        &       $i_{psf}-i_{dev}$       \\
\colsquare{colore}&{\sffamily E}        &       $i_{dev}/i_{psf}$       \\
\colsquare{colorf}&{\sffamily F}        &       $z_{model}/z_{psf}$     \\
\colsquare{colorg}&{\sffamily G}        &       $\text{\ul{$r_{psf}$}}/\text{\ul{$i_{exp}$}}$   \\
\colsquare{colorh}&{\sffamily H}        &       $i_{psf}/i_{exp}$       \\
\colsquare{colori}&{\sffamily I}        &       $\text{\ul{$g_{psf}$}}/\text{\ul{$u_{model}$}}$ \\
\colsquare{colorj}&{\sffamily J}        &       $i_{psf}/i_{model}$     \\
\colsquare{colork}&{\sffamily K}        &       $\text{\ul{$g_{psf}$}}/\text{\ul{$i_{exp}$}}$   \\
\colsquare{colorl}&{\sffamily L}        &       $\text{\ul{$r_{psf}$}}/\text{\ul{$i_{model}$}}$ \\
\colsquare{colorm}&{\sffamily M}        &       $\text{\ul{$i_{dev}$}}/\text{\ul{$i_{psf}$}}$   \\
\colsquare{colorn}&{\sffamily N}        &       $\text{\ul{$r_{psf}$}}/\text{\ul{$g_{model}$}}$ \\
\colsquare{coloro}&{\sffamily O}        &       $\text{\ul{$i_{psf}$}}/\text{\ul{$z_{model}$}}$ \\
\colsquare{colorp}&{\sffamily P}        &       $\text{\ul{$g_{petro}$}}/\text{\ul{$r_{petro}$}}$       \\
\colsquare{colorq}&{\sffamily Q}        &       $r_{psf}-r_{petro}$     \\
\colsquare{colorr}&{\sffamily R}        &       $\sqrt{\sigma_{r_{model}}^{2}+\sigma_{g_{dev}}^{2}}$    \\
\colsquare{colors}&{\sffamily S}        &       $\text{\ul{$r_{dev}$}}/\text{\ul{$r_{psf}$}}$   \\
\colsquare{colort}&{\sffamily T}        &       $\sqrt{\sigma_{r_{model}}^{2}+\sigma_{g_{exp}}^{2}}$    \\
\end{tabular}\end{minipage}
\renewcommand{\arraystretch}{1}
   \caption{Chord diagram for the experiment DR7a. Every feature is associated to a specific colour, and starting from the first features (H,J) it is possible to follow all the possible paths of the tree, depicting the different feature subsets.}\label{Fig:chorddr7a}
\end{figure*}

\begin{figure*}
\definecolor{colora}{RGB}{230,46,46}
\definecolor{colorb}{RGB}{230,83,46}
\definecolor{colorc}{RGB}{230,119,46}
\definecolor{colord}{RGB}{230,156,46}
\definecolor{colore}{RGB}{230,193,46}
\definecolor{colorf}{RGB}{230,230,46}
\definecolor{colorg}{RGB}{193,230,46}
\definecolor{colorh}{RGB}{156,230,46}
\definecolor{colori}{RGB}{119,230,46}
\definecolor{colorj}{RGB}{83,230,46}
\definecolor{colork}{RGB}{46,230,46}
\definecolor{colorl}{RGB}{46,230,83}
\definecolor{colorm}{RGB}{46,230,119}
\definecolor{colorn}{RGB}{46,230,156}
\definecolor{coloro}{RGB}{46,230,193}
\definecolor{colorp}{RGB}{46,230,230}
\definecolor{colorq}{RGB}{46,193,230}
\definecolor{colorr}{RGB}{46,156,230}
\definecolor{colors}{RGB}{46,119,230}
\definecolor{colort}{RGB}{46,83,230}
\definecolor{coloru}{RGB}{46,46,230}
\definecolor{colorv}{RGB}{83,46,230}
\definecolor{colorw}{RGB}{119,46,230}
\definecolor{colorx}{RGB}{156,46,230}
\definecolor{colory}{RGB}{193,46,230}
\definecolor{colorz}{RGB}{230,46,230}
\center{\textbf{Experiment DR7b}}\\
\begin{minipage}[]{0.60\textwidth}
\centering
\includegraphics[width=0.99\textwidth]{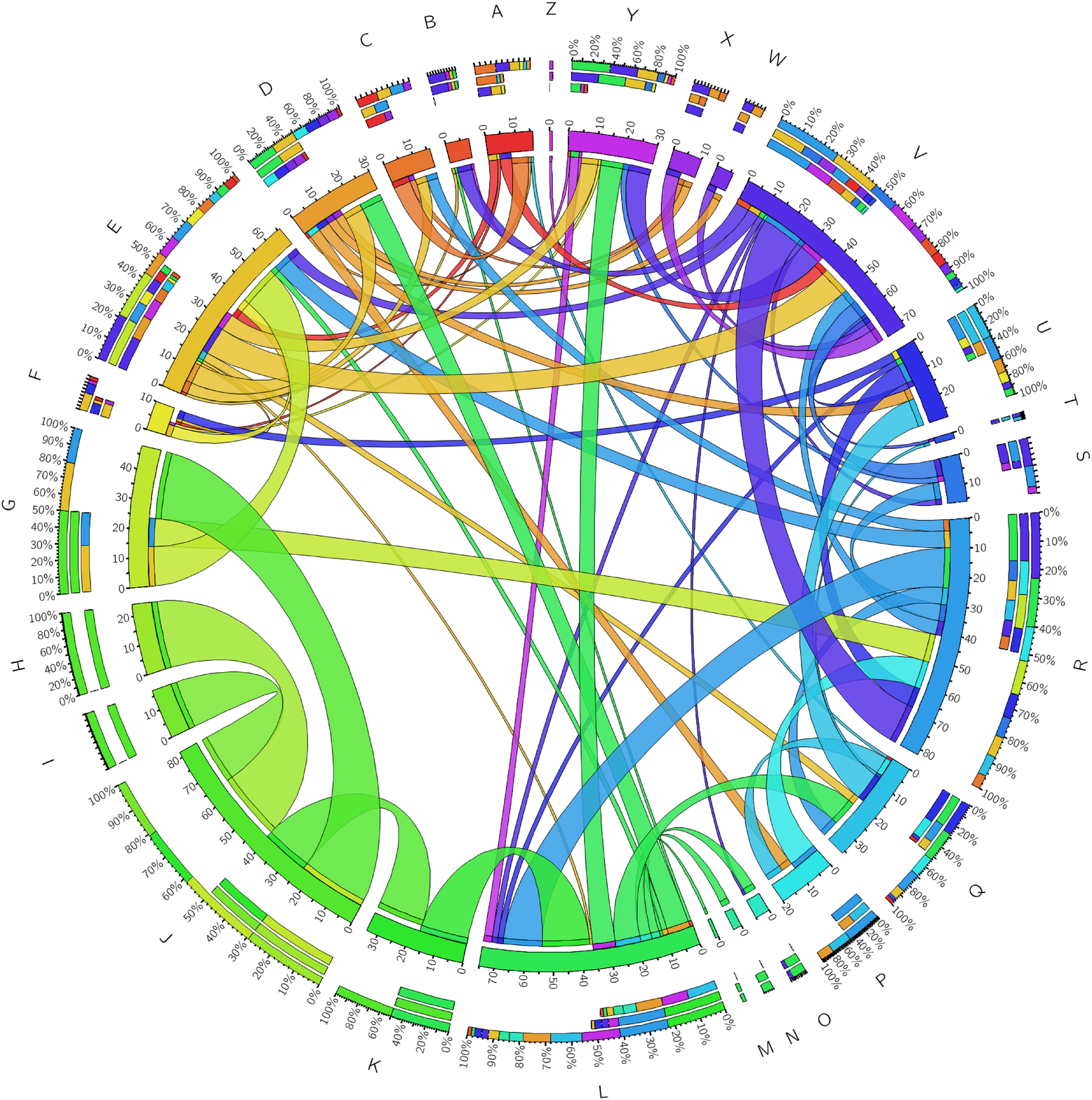}
\end{minipage}\hfill
\begin{minipage}[]{0.395\textwidth}
\renewcommand{\arraystretch}{1.5}
\tiny
\begin{tabular}{cccccc}
 & name & feature &  & name & feature\\
\colsquare{colora}&{\sffamily A}        &       $\text{\ul{$r_{psf}$}}/\text{\ul{$g_{exp}$}}$ & \colsquare{colorn}&{\sffamily N}        &       $i_{psf}/z_{petro}$\\
\colsquare{colorb}&{\sffamily B}        &       $g_{psf}-g_{dev}$ & \colsquare{coloro}&{\sffamily O}        &       $r_{petroR}-z_{petroR90}$\\ 
\colsquare{colorc}&{\sffamily C}        &       $\text{\ul{$z_{psf}$}}/\text{\ul{$i_{exp}$}}$ & \colsquare{colorp}&{\sffamily P}        &       $\text{\ul{$g_{psf}$}}/\text{\ul{$r_{exp}$}}$\\
\colsquare{colord}&{\sffamily D}        &       $\text{\ul{$z_{psf}$}}/\text{\ul{$i_{model}$}}$ & \colsquare{colorq}&{\sffamily Q}        &       $\text{\ul{$i_{psf}$}}/\text{\ul{$z_{model}$}}$\\
\colsquare{colore}&{\sffamily E}        &       $\text{\ul{$i_{dev}$}}/\text{\ul{$i_{psf}$}}$ & \colsquare{colorr}&{\sffamily R}        &       $i_{psf}-i_{petro}$\\
\colsquare{colorf}&{\sffamily F}        &       $\text{\ul{$r_{dev}$}}/\text{\ul{$r_{psf}$}}$ & \colsquare{colors}&{\sffamily S}        &       $i_{psf}-i_{dev}$\\
\colsquare{colorg}&{\sffamily G}        &       $\text{\ul{$r_{psf}$}}/\text{\ul{$i_{model}$}}$ & \colsquare{colort}&{\sffamily T}        &       $\text{\ul{$g_{dev}$}}/\text{\ul{$r_{petro}$}}$\\
\colsquare{colorh}&{\sffamily H}        &       $i_{psf}/i_{model}$ & \colsquare{coloru}&{\sffamily U}        &       $\text{\ul{$g_{psf}$}}/\text{\ul{$r_{model}$}}$\\
\colsquare{colori}&{\sffamily I}        &       $i_{psf}/i_{exp}$ & \colsquare{colorv}&{\sffamily V}        &       $r_{psf}-r_{petro}$\\
\colsquare{colorj}&{\sffamily J}        &       $\text{\ul{$g_{psf}$}}/\text{\ul{$u_{model}$}}$ & \colsquare{colorw}&{\sffamily W}        &       $\text{\ul{$r_{psf}$}}/\text{\ul{$g_{model}$}}$\\
\colsquare{colork}&{\sffamily K}        &       $\text{\ul{$r_{psf}$}}/\text{\ul{$i_{exp}$}}$ & \colsquare{colorx}&{\sffamily X}        &       $\text{\ul{$r_{psf}$}}/\text{\ul{$g_{petro}$}}$\\
\colsquare{colorl}&{\sffamily L}        &       $\text{\ul{$z_{model}$}}/\text{\ul{$z_{psf}$}}$ & \colsquare{colory}&{\sffamily Y}        &       $\sqrt{\sigma_{g_{model}}^{2}+\sigma_{g_{dev}}^{2}}$\\
\colsquare{colorm}&{\sffamily M}        &       $g_{model}-g_{exp}$ & \colsquare{colorz}&{\sffamily Z}        &       $z_{exp}/z_{psf}$\\
 
\end{tabular}\end{minipage}
\renewcommand{\arraystretch}{1}
   \caption{Chord diagram for the experiment DR7b. Every feature is associated to a specific colour, and starting from the first features (H,I) it is possible to follow all the possible paths of the tree, depicting the different feature subsets.}\label{Fig:chorddr7b}
\end{figure*}

\begin{table*}\resizebox{0.88\textwidth}{!}{\centering
\definecolor{colora}{RGB}{230,46,46}
\definecolor{colorb}{RGB}{230,83,46}
\definecolor{colorc}{RGB}{230,119,46}
\definecolor{colord}{RGB}{230,156,46}
\definecolor{colore}{RGB}{230,193,46}
\definecolor{colorf}{RGB}{230,230,46}
\definecolor{colorg}{RGB}{193,230,46}
\definecolor{colorh}{RGB}{156,230,46}
\definecolor{colori}{RGB}{119,230,46}
\definecolor{colorj}{RGB}{83,230,46}
\definecolor{colork}{RGB}{46,230,46}
\definecolor{colorl}{RGB}{46,230,83}
\definecolor{colorm}{RGB}{46,230,119}
\definecolor{colorn}{RGB}{46,230,156}
\definecolor{coloro}{RGB}{46,230,193}
\definecolor{colorp}{RGB}{46,230,230}
\definecolor{colorq}{RGB}{46,193,230}
\definecolor{colorr}{RGB}{46,156,230}
\definecolor{colors}{RGB}{46,119,230}
\definecolor{colort}{RGB}{46,83,230}
\definecolor{coloru}{RGB}{46,46,230}
\definecolor{colorv}{RGB}{83,46,230}
\definecolor{colorw}{RGB}{119,46,230}
\definecolor{colorx}{RGB}{156,46,230}
\definecolor{colory}{RGB}{193,46,230}
\definecolor{colorz}{RGB}{230,46,230}

\begin{tikzpicture}
\matrix (m)[matrix of nodes, style={nodes={rectangle,minimum width=8em,  row sep=-\pgflinewidth, column sep=-\pgflinewidth}}, ampersand replacement =\&]
{
  {\bf id} \&
  {\bf Feature 1} \& {\bf Feature 2} \& {\bf  Feature 3} \&  {\bf Feature 4} \&  {\bf Feature 5} \&  {\bf Feature 6} \&  {\bf Feature 7} \&  {\bf Feature 8} \& {\bf  Feature 9} \& {\bf  Feature 10} \\
\hline
$1$     \&      \raisebox{1.2em}{$$} \&$        \raisebox{1.2em}{$$}    $\&     \raisebox{1.2em}{$$}    \&      \raisebox{1.2em}{$$}    \&      \raisebox{1.2em}{$$}    \&      \raisebox{1.2em}{$$}    \&      \raisebox{1.2em}{$$}    \&      \raisebox{1.2em}{$$}\colorbox{white}{$r_{psf}-r_{petro}$}       \&      \raisebox{1.2em}{$$}\raisebox{-1.2em}{$$}\colorbox{white}{$i_{psf}-i_{dev}$}    \&       \raisebox{1.2em}{$$}    \\
$2$     \&      $$      \&      $$      \&      $$      \&      $ $     \&      $ $       \&      \colorbox{white}{${\text{\ul{$g_{psf}$}}/\text{\ul{$r_{model}$}}}$}     \&      $ $       \&      \raisebox{-1.4em}{$$}   \&      \raisebox{1.2em}{$$}\raisebox{-1.4em}{$$}\colorbox{white}{${\text{\ul{$i_{dev}$}}/\text{\ul{$i_{psf}$}}}$}      \&      $ $       \\
$3$     \&      $$      \&      $$      \&      $$      \&      $ $     \&      $ $       \&      \raisebox{-1.2em}{$$}   \&      $ $     \&      \raisebox{1.4em}{$$}\raisebox{-1.2em}{$$}\colorbox{white}{$i_{psf}-i_{dev}$}    \&      \raisebox{1.4em}{$$}\raisebox{-1.2em}{$$}\colorbox{white}{$r_{psf}-r_{petro}$}  \&      \colorbox{white}{$\sqrt{\sigma_{g_{model}}^2+\sigma_{g_{dev}}^2}$}      \\
$4$     \&      $$      \&      $$      \&      $$      \&      $ $     \&      $ $       \&      \raisebox{1.2em}{$$}    \&      $ $     \&      \raisebox{1.2em}{$$}    \&      \raisebox{1.2em}{$$}\raisebox{-1.2em}{$$}\colorbox{white}{$i_{psf}-i_{dev}$}    \&      $ $       \\
$5$     \&      $$      \&      $$      \&      $$      \&      $ $     \&      \colorbox{white}{${\text{\ul{$i_{psf}$}}/\text{\ul{$z_{model}$}}}$}     \&      $ $       \&      $ $     \&      \colorbox{white}{$r_{psf}-r_{petro}$}   \&      \raisebox{1.2em}{$$}\colorbox{white}{${\text{\ul{$i_{dev}$}}/\text{\ul{$i_{psf}$}}}$}   \&      \raisebox{-1.2em}{$$}   \\
$6$     \&      $$      \&      $$      \&      $$      \&      $ $     \&      $ $       \&      \colorbox{white}{${\text{\ul{$g_{psf}$}}/\text{\ul{$r_{exp}$}}}$}       \&      $ $       \&      \raisebox{-1.4em}{$$}   \&      \raisebox{-1.2em}{$$}   \&      \raisebox{1.2em}{$$}\raisebox{-1.4em}{$$}\colorbox{white}{$g_{psf}-g_{dev}$}    \\
$7$     \&      $$      \&      $$      \&      $$      \&      $ $     \&      $ $       \&      $ $     \&      $ $     \&      \raisebox{1.4em}{$$}\raisebox{-1.2em}{$$}\colorbox{white}{$i_{psf}-i_{dev}$}    \&      \raisebox{1.2em}{$$}    \&      \raisebox{1.4em}{$$}\colorbox{white}{$\sqrt{\sigma_{g_{model}}^2+\sigma_{g_{dev}}^2}$}  \\
$8$     \&      $$      \&      $$      \&      $$      \&      $ $     \&      $ $       \&      $ $     \&      $ $     \&      \raisebox{1.2em}{$$}\colorbox{white}{${\text{\ul{$i_{dev}$}}/\text{\ul{$i_{psf}$}}}$}   \&      $ $       \&      \raisebox{-1.2em}{$$}   \\
$9$     \&      \colorbox{white}{$i_{psf}/i_{exp}$}     \&      $$      \&      \colorbox{white}{${\text{\ul{$r_{psf}$}}/\text{\ul{$i_{exp}$}}}$}       \&      \colorbox{white}{${\text{\ul{$z_{model}$}}/\text{\ul{$z_{psf}$}}}$}     \&      \raisebox{-1.4em}{$$}   \&      \raisebox{-1.4em}{$$}   \&      \colorbox{white}{$i_{psf}-i_{petro}$}   \&      \raisebox{-1.4em}{$$}   \&      $ $       \&      \raisebox{1.4em}{$$}\raisebox{-1.4em}{$$}\colorbox{white}{$g_{psf}-g_{dev}$}    \\
$10$    \&      $$      \&      $$      \&      $$      \&      $ $     \&      \raisebox{1.4em}{$$}    \&      \raisebox{1.4em}{$$}    \&      $ $       \&      \raisebox{1.4em}{$$}\colorbox{white}{$i_{psf}-i_{dev}$} \&      $ $       \&      \raisebox{1.4em}{$$}\raisebox{-1.2em}{$$}\colorbox{white}{$\sqrt{\sigma_{g_{model}}^2+\sigma_{g_{dev}}^2}$}     \\
$11$    \&      $$ \&   $$      \&      $$      \&      $ $     \&      $ $       \&      \colorbox{white}{${\text{\ul{$g_{psf}$}}/\text{\ul{$r_{model}$}}}$}     \&      $ $       \&      \raisebox{-1.4em}{$$}   \&      $ $     \&      \raisebox{1.2em}{$$}\raisebox{-1.4em}{$$}\colorbox{white}{$r_{petroR}-z_{petroR90}$}    \\
$12$    \&      $$      \&      $$      \&      $$      \&      $ $     \&      $ $       \&      $ $     \&      $ $     \&      \raisebox{1.4em}{$$}\colorbox{white}{${\text{\ul{$i_{dev}$}}/\text{\ul{$i_{psf}$}}}$}   \&      \colorbox{white}{$r_{psf}-r_{petro}$}   \&      \raisebox{1.4em}{$$}\raisebox{-1.2em}{$$}\colorbox{white}{$\sqrt{\sigma_{g_{model}}^2+\sigma_{g_{dev}}^2}$}     \\
$13*$   \&      $$      \&      $$      \&      $$      \&      $ $     \&      \colorbox{white}{${\text{\ul{$z_{psf}$}}/\text{\ul{$i_{model}$}}}$}     \&      \raisebox{-1.4em}{$$}   \&      $ $       \&      \raisebox{-1.4em}{$$}   \&      $ $     \&      \raisebox{1.2em}{$$}\raisebox{-1.4em}{$$}\colorbox{white}{$g_{psf}-g_{dev}$}    \\
$14$    \&      $$      \&      $$      \&      $$      \&      $ $     \&      $$      \&      \raisebox{1.4em}{$$}    \&      $ $       \&      \raisebox{1.4em}{$$}\colorbox{white}{$i_{psf}-i_{dev}$} \&      $ $       \&      \raisebox{1.4em}{$$}\raisebox{-1.2em}{$$}\colorbox{white}{$\sqrt{\sigma_{g_{model}}^2+\sigma_{g_{dev}}^2}$}     \\
$15$    \&      $$      \&      $$      \&      $$      \&      $ $     \&      $ $       \&      \colorbox{white}{${\text{\ul{$g_{psf}$}}/\text{\ul{$r_{exp}$}}}$}       \&      $ $       \&      \raisebox{-1.4em}{$$}   \&      $ $     \&      \raisebox{1.2em}{$$}\raisebox{-1.4em}{$$}\colorbox{white}{$g_{psf}-g_{dev}$}    \\
$16$    \&      $$      \&      $$      \&      $$      \&      $ $     \&      $ $       \&      $ $     \&      $ $     \&      \raisebox{1.4em}{$$}    \&      $ $       \&      \raisebox{1.4em}{$$}\raisebox{-1.2em}{$$}\colorbox{white}{$\sqrt{\sigma_{g_{model}}^2+\sigma_{g_{dev}}^2}$}     \\
$17$    \&      \raisebox{-1.2em}{$$}   \&      $$      \&      \raisebox{-1.2em}{$$}   \&      \raisebox{-1.2em}{$$}   \&      \raisebox{-1.2em}{$$}   \&      \raisebox{-1.2em}{$$}   \&      \raisebox{-1.2em}{$$}   \&      $ $       \&      \raisebox{-1.2em}{$$}   \&      \raisebox{1.2em}{$$}\raisebox{-1.2em}{$$}\colorbox{white}{$g_{psf}-g_{dev}$}    \\
$18$    \&      \raisebox{1.2em}{$$}    \&      $$      \&      \raisebox{1.2em}{$$}    \&      \raisebox{1.2em}{$$}    \&      \raisebox{1.2em}{$$}    \&      \raisebox{1.2em}{$$}    \&      \raisebox{1.2em}{$$}    \&      $ $       \&      \raisebox{1.2em}{$$}    \&      \raisebox{1.2em}{$$}\raisebox{-1.4em}{$$}\colorbox{white}{$z_{exp}/z_{psf}$}    \\
$19$    \&      $$      \&      $$      \&      $$      \&      $ $     \&      $ $       \&      $ $     \&      \colorbox{white}{${\text{\ul{$r_{dev}$}}/\text{\ul{$r_{psf}$}}}$}       \&      \colorbox{white}{${\text{\ul{$i_{dev}$}}/\text{\ul{$i_{psf}$}}}$}       \&      \colorbox{white}{$\sqrt{\sigma_{g_{model}}^2+\sigma_{g_{dev}}^2}$}      \&      \raisebox{1.4em}{$$}\raisebox{-1.2em}{$$}\colorbox{white}{${\text{\ul{$z_{model}$}}/\text{\ul{$z_{psf}$}}}$}    \\
$20$    \&      $$      \&      \colorbox{white}{\text{\ul{$g_{psf}$}}/\text{\ul{$u_{model}$}}} \&      $$      \&      $ $       \&      \colorbox{white}{${\text{\ul{$i_{psf}$}}/\text{\ul{$z_{model}$}}}$}     \&      \colorbox{white}{${\text{\ul{$g_{psf}$}}/\text{\ul{$r_{model}$}}}$}     \&      \raisebox{-1.4em}{$$}   \&      $ $       \&      \raisebox{-1.4em}{$$}   \&      \raisebox{1.2em}{$$}\raisebox{-1.4em}{$$}\colorbox{white}{$g_{psf}-g_{dev}$}    \\
$21$    \&      $$      \&      $$      \&      $$      \&      $$      \&      $ $       \&      \raisebox{-1.2em}{$$}   \&      \raisebox{1.4em}{$$}\colorbox{white}{${\text{\ul{$z_{model}$}}/\text{\ul{$z_{psf}$}}}$} \&      $ $       \&      \raisebox{1.4em}{$$}\colorbox{white}{$r_{psf}-r_{petro}$}       \&      \raisebox{1.4em}{$$}\raisebox{-1.2em}{$$}\colorbox{white}{$\sqrt{\sigma_{g_{model}}^2+\sigma_{g_{dev}}^2}$}     \\
$22$    \&      $$      \&      $$      \&      $$      \&      \colorbox{white}{$i_{psf}-i_{petro}$}   \&      $ $       \&      \raisebox{-1.4em}{$$}   \&      \raisebox{-1.4em}{$$}   \&      $ $       \&      \raisebox{-1.4em}{$$}   \&      \raisebox{1.2em}{$$}\raisebox{-1.4em}{$$}\colorbox{white}{$g_{psf}-g_{dev}$}    \\
$23$    \&      $$      \&      $$      \&      $$      \&      $$      \&      \raisebox{-1.2em}{$$}   \&      \raisebox{1.4em}{$$}\raisebox{-1.2em}{$$}\colorbox{white}{${\text{\ul{$g_{dev}$}}/\text{\ul{$r_{petro}$}}}$}    \&      \raisebox{1.4em}{$$}\raisebox{-1.2em}{$$}\colorbox{white}{$r_{psf}-r_{petro}$}  \&      \raisebox{-1.2em}{$$}   \&      \raisebox{1.4em}{$$}    \&      \raisebox{1.4em}{$$}\colorbox{white}{$\sqrt{\sigma_{g_{model}}^2+\sigma_{g_{dev}}^2}$}  \\
$24$    \&      $$      \&      $$      \&      $$      \&      $ $     \&      \raisebox{1.2em}{$$}    \&      \raisebox{1.2em}{$$}    \&      \raisebox{1.2em}{$$}    \&      \raisebox{1.2em}{$$}\colorbox{white}{$r_{psf}-r_{petro}$}       \&      \colorbox{white}{${\text{\ul{$z_{model}$}}/\text{\ul{$z_{psf}$}}}$}     \&      \raisebox{-1.2em}{$$}   \\
$25$    \&      $$      \&      $$      \&      $$      \&      $ $     \&      $ $       \&      $ $     \&      \colorbox{white}{${\text{\ul{$i_{dev}$}}/\text{\ul{$i_{psf}$}}}$}       \&      \raisebox{-1.4em}{$$}   \&      \raisebox{-1.4em}{$$}   \&      \raisebox{1.2em}{$$}\raisebox{-1.4em}{$$}\colorbox{white}{$r_{petroR}-z_{petroR90}$}    \\
$26$    \&      $$      \&      $$      \&      $$      \&      $ $     \&      $ $       \&      \colorbox{white}{${\text{\ul{$r_{psf}$}}/\text{\ul{$g_{exp}$}}}$}       \&      \raisebox{-1.2em}{$$}   \&      \raisebox{1.4em}{$$}\raisebox{-1.2em}{$$}\colorbox{white}{${\text{\ul{$r_{dev}$}}/\text{\ul{$r_{psf}$}}}$}      \&      \raisebox{1.4em}{$$}\colorbox{white}{$\sqrt{\sigma_{g_{model}}^2+\sigma_{g_{dev}}^2}$}  \&      \raisebox{1.4em}{$$}\colorbox{white}{${\text{\ul{$z_{model}$}}/\text{\ul{$z_{psf}$}}}$} \\
$27$    \&      $$      \&      $$      \&      $$      \&      \raisebox{-1.2em}{$$}   \&      \colorbox{white}{${\text{\ul{$z_{psf}$}}/\text{\ul{$i_{exp}$}}}$}       \&      $ $       \&      \raisebox{1.2em}{$$}\raisebox{-1.2em}{$$}\colorbox{white}{${\text{\ul{$r_{dev}$}}/\text{\ul{$r_{psf}$}}}$}      \&      \raisebox{1.2em}{$$}\raisebox{-1.2em}{$$}\colorbox{white}{${\text{\ul{$i_{dev}$}}/\text{\ul{$i_{psf}$}}}$}      \&      \raisebox{-1.2em}{$$}   \&      \raisebox{-1.2em}{$$}   \\
$28$    \&      $$      \&      $$      \&      $$      \&      \raisebox{1.2em}{$$}    \&      $ $       \&      $ $     \&      \raisebox{1.2em}{$$}    \&      \raisebox{1.2em}{$$}    \&      \raisebox{1.2em}{$$}    \&      \raisebox{1.2em}{$$}\raisebox{-1.2em}{$$}\colorbox{white}{$r_{petroR}-z_{petroR90}$}    \\
$29$    \&      \colorbox{white}{$i_{psf}/i_{model}$}   \&      $$      \&      \colorbox{white}{$\text{\ul{$r_{psf}$}}/\text{\ul{$i_{model}$}}$}       \&      $ $       \&      $ $     \&      \raisebox{-1.4em}{$$}   \&      $ $     \&      $ $       \&      $ $     \&      \raisebox{1.2em}{$$}\raisebox{-1.4em}{$$}\colorbox{white}{$i_{psf}/z_{petro}$}  \\
$30$    \&      $$      \&      $$      \&      $$      \&      $ $     \&      $ $       \&      \raisebox{1.4em}{$$}\colorbox{white}{${\text{\ul{$r_{psf}$}}/\text{\ul{$g_{petro}$}}}$} \&      $ $       \&      $ $     \&      $ $     \&      \raisebox{1.4em}{$$}\raisebox{-1.2em}{$$}\colorbox{white}{$\sqrt{\sigma_{g_{model}}^2+\sigma_{g_{dev}}^2}$}     \\
$31$    \&      $$      \&      $$      \&      $$      \&      $ $     \&      \raisebox{-1.4em}{$$}   \&      \raisebox{-1.4em}{$$}   \&      $ $       \&      $ $     \&      $ $     \&      \raisebox{1.2em}{$$}\raisebox{-1.4em}{$$}\colorbox{white}{$i_{psf}/z_{petro}$}  \\
$32$    \&      $$      \&      $$      \&      $$      \&      $ $     \&      \raisebox{1.4em}{$$}    \&      \raisebox{1.4em}{$$}\raisebox{-1.2em}{$$}\colorbox{white}{${\text{\ul{$r_{psf}$}}/\text{\ul{$g_{exp}$}}}$}      \&      $ $       \&      $ $     \&      $ $     \&      \raisebox{1.4em}{$$}\colorbox{white}{$\sqrt{\sigma_{g_{model}}^2+\sigma_{g_{dev}}^2}$}  \\
$33$    \&      $$      \&      $$      \&      $$      \&      $ $     \&      \colorbox{white}{${\text{\ul{$i_{psf}$}}/\text{\ul{$z_{model}$}}}$}     \&      \raisebox{1.2em}{$$}\colorbox{white}{${\text{\ul{$g_{psf}$}}/\text{\ul{$r_{model}$}}}$} \&      $ $       \&      $ $     \&      $ $     \&      \raisebox{-1.2em}{$$}   \\
$34$    \&      $$      \&      $$      \&      $$      \&      \colorbox{white}{${\text{\ul{$i_{dev}$}}/\text{\ul{$i_{psf}$}}}$}       \&      \raisebox{-1.4em}{$$}   \&      \raisebox{-1.4em}{$$}   \&      \colorbox{white}{$r_{psf}-r_{petro}$}   \&      \colorbox{white}{$i_{psf}-i_{petro}$}   \&      \colorbox{white}{${\text{\ul{$z_{model}$}}/\text{\ul{$z_{psf}$}}}$}     \&      \raisebox{1.2em}{$$}\raisebox{-1.4em}{$$}\colorbox{white}{$g_{psf}-g_{dev}$}    \\
$35$    \&      $$      \&      $$      \&      $$      \&      $$      \&      \raisebox{1.4em}{$$}    \&      \raisebox{1.4em}{$$}    \&      $ $       \&      $ $     \&      $ $     \&      \raisebox{1.4em}{$$}\colorbox{white}{$\sqrt{\sigma_{g_{model}}^2+\sigma_{g_{dev}}^2}$}  \\
$36$    \&      $$      \&      $$      \&      $$      \&      $ $     \&      $ $       \&      \colorbox{white}{${\text{\ul{$r_{psf}$}}/\text{\ul{$g_{exp}$}}}$}       \&      $ $       \&      $ $     \&      $ $     \&      \raisebox{-1.2em}{$$}   \\
$37$    \&      $$      \&      $$      \&      $$      \&      $ $     \&      $ $       \&      $ $     \&      $ $     \&      $ $     \&      $ $     \&      \raisebox{1.2em}{$$}\raisebox{-1.2em}{$$}\colorbox{white}{$i_{psf}/z_{petro}$}  \\
$38$    \&      $$      \&      $$      \&      $$      \&      $ $     \&      \colorbox{white}{${\text{\ul{$z_{psf}$}}/\text{\ul{$i_{model}$}}}$}     \&      \raisebox{-1.4em}{$$}   \&      $ $       \&      $ $     \&      $ $     \&      \raisebox{1.2em}{$$}\raisebox{-1.4em}{$$}\colorbox{white}{$r_{petroR}-z_{petroR90}$}    \\
$39$    \&      $$      \&      $$      \&      $$      \&      $ $     \&      $ $       \&      \raisebox{1.4em}{$$}    \&      $ $     \&      $ $     \&      $ $       \&      \raisebox{1.4em}{$$}\raisebox{-1.2em}{$$}\colorbox{white}{$\sqrt{\sigma_{g_{model}}^2+\sigma_{g_{dev}}^2}$}     \\
$40$    \&      $$      \&      $$      \&      $$      \&      $ $     \&      $ $       \&      \colorbox{white}{${\text{\ul{$r_{psf}$}}/\text{\ul{$g_{model}$}}}$}     \&      $ $       \&      $ $     \&      $ $     \&      \raisebox{1.2em}{$$}\raisebox{-1.2em}{$$}\colorbox{white}{$r_{petroR}-z_{petroR90}$}    \\
$41$    \&      \raisebox{-1.2em}{$$}   \&      \raisebox{-1.2em}{$$}   \&      \raisebox{-1.2em}{$$}   \&      \raisebox{-1.2em}{$$} \&      \raisebox{-1.2em}{$$}   \&      \raisebox{-1.2em}{$$}   \&      \raisebox{-1.2em}{$$}   \&      \raisebox{-1.2em}{$$}   \&      \raisebox{-1.2em}{$$}   \&      \raisebox{-1.2em}{$$} \colorbox{white}{$g_{model}-g_{exp}$}   \\};

\begin{pgfonlayer}{myback}
\fillpattern[dots][colori]{m-2-2}{m-18-2}
\fillpattern[dots][colorh]{m-19-2}{m-42-2}
\fillpattern[vertical lines][colorj]{m-2-3}{m-42-3}
\fillpattern[vertical lines][colork]{m-2-4}{m-18-4}
\fillpattern[vertical lines][colorg]{m-19-4}{m-42-4}
\fillpattern[dots][colorl]{m-2-5}{m-18-5}
\fillpattern[horizontal lines][colorr]{m-19-5}{m-28-5}
\fillpattern[dots][colore]{m-29-5}{m-42-5}
\fillpattern[vertical lines][colorq]{m-2-6}{m-10-6}
\fillpattern[vertical lines][colord]{m-11-6}{m-18-6}
\fillpattern[vertical lines][colorq]{m-19-6}{m-24-6}
\fillpattern[vertical lines][colorc]{m-25-6}{m-32-6}
\fillpattern[vertical lines][colorq]{m-33-6}{m-35-6}
\fillpattern[vertical lines][colord]{m-36-6}{m-42-6}
\fillpattern[vertical lines][coloru]{m-2-7}{m-4-7}
\fillpattern[vertical lines][colorp]{m-5-7}{m-10-7}
\fillpattern[vertical lines][coloru]{m-11-7}{m-14-7}
\fillpattern[vertical lines][colorp]{m-15-7}{m-18-7}
\fillpattern[vertical lines][coloru]{m-19-7}{m-23-7}
\fillpattern[vertical lines][colort]{m-24-7}{m-24-7}
\fillpattern[vertical lines][colora]{m-25-7}{m-30-7}
\fillpattern[vertical lines][colorx]{m-31-7}{m-32-7}
\fillpattern[vertical lines][colora]{m-33-7}{m-33-7}
\fillpattern[vertical lines][coloru]{m-34-7}{m-35-7}
\fillpattern[vertical lines][colora]{m-36-7}{m-39-7}
\fillpattern[vertical lines][colorw]{m-40-7}{m-42-7}
\fillpattern[horizontal lines][colorr]{m-2-8}{m-18-8}
\fillpattern[dots][colorf]{m-19-8}{m-21-8}
\fillpattern[dots][colorl]{m-22-8}{m-23-8}
\fillpattern[horizontal lines][colorv]{m-24-8}{m-24-8}
\fillpattern[dots][colore]{m-25-8}{m-27-8}
\fillpattern[dots][colorf]{m-28-8}{m-28-8}
\fillpattern[horizontal lines][colorv]{m-29-8}{m-42-8}
\fillpattern[horizontal lines][colorv]{m-2-9}{m-3-9}
\fillpattern[horizontal lines][colors]{m-4-9}{m-4-9}
\fillpattern[horizontal lines][colorv]{m-5-9}{m-7-9}
\fillpattern[horizontal lines][colors]{m-8-9}{m-8-9}
\fillpattern[dots][colore]{m-9-9}{m-10-9}
\fillpattern[horizontal lines][colors]{m-11-9}{m-12-9}
\fillpattern[dots][colore]{m-13-9}{m-14-9}
\fillpattern[horizontal lines][colors]{m-15-9}{m-16-9}
\fillpattern[dots][colore]{m-17-9}{m-24-9}
\fillpattern[horizontal lines][colorv]{m-25-9}{m-26-9}
\fillpattern[dots][colorf]{m-27-9}{m-27-9}
\fillpattern[dots][colore]{m-28-9}{m-28-9}
\fillpattern[horizontal lines][colorr]{m-29-9}{m-42-9}
\fillpattern[dots][colors]{m-2-10}{m-2-10}
\fillpattern[dots][colore]{m-3-10}{m-3-10}
\fillpattern[horizontal lines][colorv]{m-4-10}{m-4-10}
\fillpattern[horizontal lines][colors]{m-5-10}{m-5-10}
\fillpattern[horizontal lines][colore]{m-6-10}{m-7-10}
\fillpattern[horizontal lines][colorv]{m-8-10}{m-18-10}
\fillpattern[north west lines][colory]{m-19-10}{m-21-10}
\fillpattern[horizontal lines][colorv]{m-22-10}{m-23-10}
\fillpattern[dots][colorl]{m-24-10}{m-26-10}
\fillpattern[north west lines][colory]{m-27-10}{m-28-10}
\fillpattern[dots][colorl]{m-29-10}{m-42-10}
\fillpattern[north west lines][colory]{m-2-11}{m-6-11}
\fillpattern[horizontal lines][colorb]{m-7-11}{m-7-11}
\fillpattern[north west lines][colory]{m-8-11}{m-9-11}
\fillpattern[horizontal lines][colorb]{m-10-11}{m-10-11}
\fillpattern[north west lines][colory]{m-11-11}{m-11-11}
\fillpattern[grid][coloro]{m-12-11}{m-12-11}
\fillpattern[north west lines][colory]{m-13-11}{m-13-11}
\fillpattern[horizontal lines][colorb]{m-14-11}{m-14-11}
\fillpattern[north west lines][colory]{m-15-11}{m-15-11}
\fillpattern[horizontal lines][colorb]{m-16-11}{m-16-11}
\fillpattern[north west lines][colory]{m-17-11}{m-17-11}
\fillpattern[horizontal lines][colorb]{m-18-11}{m-18-11}
\fillpattern[dots][colorz]{m-19-11}{m-19-11}
\fillpattern[dots][colorl]{m-20-11}{m-20-11}
\fillpattern[horizontal lines][colorb]{m-21-11}{m-21-11}
\fillpattern[north west lines][colory]{m-22-11}{m-22-11}
\fillpattern[horizontal lines][colorb]{m-23-11}{m-23-11}
\fillpattern[north west lines][colory]{m-24-11}{m-25-11}
\fillpattern[grid][coloro]{m-26-11}{m-26-11}
\fillpattern[dots][colorl]{m-27-11}{m-28-11}
\fillpattern[grid][coloro]{m-29-11}{m-29-11}
\fillpattern[vertical lines][colorn]{m-30-11}{m-30-11}
\fillpattern[north west lines][colory]{m-31-11}{m-31-11}
\fillpattern[vertical lines][colorn]{m-32-11}{m-32-11}
\fillpattern[north west lines][colory]{m-33-11}{m-34-11}
\fillpattern[horizontal lines][colorb]{m-35-11}{m-35-11}
\fillpattern[north west lines][colory]{m-36-11}{m-37-11}
\fillpattern[vertical lines][colorn]{m-38-11}{m-38-11}
\fillpattern[grid][coloro]{m-39-11}{m-39-11}
\fillpattern[north west lines][colory]{m-40-11}{m-40-11}
\fillpattern[grid][coloro]{m-41-11}{m-41-11}
\fillpattern[horizontal lines][colorm]{m-42-11}{m-42-11}
\end{pgfonlayer}
\end{tikzpicture}
}
\caption{Detailed feature branches obtained from the feature selection for the experiment DR7b. The $13th$ branch, indicated with the $*$ symbol, is the best performing subset with respect to the experiments using the RF. The \emph{ratios} and \emph{photometric ratios} are indicated, respectively, with vertical lines and dots. The \emph{differences} are marked with horizontal lines and the \emph{errors} are with north west lines. Finally, the only feature composed by radius is indicated with a grid. The color code for the features is the as same shown in the chord diagram in Fig.~\ref{Fig:chorddr7b}}\label{tab:treedefinition_dr7b}\end{table*}

\clearpage
\onecolumn

\section{SDSS QSO query}\label{queryQSO}

In the following, the statements used to query the SDSS database are provided.
\subsection{Experiment DR7}
\begin{verbatim}
SELECT  
    s.specObjID, p.objid, p.ra, p.dec, s.targetObjID, s.z, s.zErr, 
    p.psfMag_u, p.psfMag_g, p.psfMag_r, p.psfMag_i, p.psfMag_z, 
    p.psfMagErr_u, p.psfMagErr_g, p.psfMagErr_r, p.psfMagErr_i, p.psfMagErr_z, 
    p.modelMag_u, p.modelMag_g, p.modelMag_r, p.modelMag_i, p.modelMag_z, 
    p.modelMagErr_u, p.modelMagErr_g, p.modelMagErr_r, p.modelMagErr_i, p.modelMagErr_z, 
    p.devMag_u, p.devMag_g, p.devMag_r, p.devMag_i, p.devMag_z, 
    p.devMagErr_u, p.devMagErr_g, p.devMagErr_r, p.devMagErr_i, p.devMagErr_z, 
    p.expMag_u, p.expMag_g, p.expMag_r, p.expMag_i, p.expMag_z, 
    p.expMagErr_u, p.expMagErr_g, p.expMagErr_r, p.expMagErr_i, p.expMagErr_z, 
    p.petroMag_u, p.petroMag_g, p.petroMag_r, p.petroMag_i, p.petroMag_z, 
    p.petroMagErr_u, p.petroMagErr_g, p.petroMagErr_r, p.petroMagErr_i, p.petroMagErr_z, 
    p.extinction_u, p.extinction_g, p.extinction_r, p.extinction_i, p.extinction_z, 
    p.devRad_u, p.devRad_g, p.devRad_r, p.devRad_i, p.devRad_z, 
    p.expRad_u, p.expRad_g, p.expRad_r, p.expRad_i, p.expRad_z, 
    p.petroRad_u, p.petroRad_g, p.petroRad_r, p.petroRad_i, p.petroRad_z, 
    p.petroR90_u, p.petroR90_g, p.petroR90_r, p.petroR90_i, p.petroR90_z, 
    p.petroR50_u, p.petroR50_g, p.petroR50_r, p.petroR50_i, p.petroR50_z, 
    p.devAB_u, p.devAB_g, p.devAB_r, p.devAB_i, p.devAB_z, 
    p.expAB_u, p.expAB_g, p.expAB_r, p.expAB_i, p.expAB_z i

FROM 
    SpecPhoto as s, PhotoObjAll as p 

WHERE 
    p.mode = 1 AND p.SpecObjID = s.SpecObjID AND
    dbo.fPhotoFlags('PEAKCENTER') != 0 AND
    dbo.fPhotoFlags('NOTCHECKED') != 0 AND
    dbo.fPhotoFlags('DEBLEND_NOPEAK') != 0 AND
    dbo.fPhotoFlags('PSF_FLUX_INTERP') != 0 AND
    dbo.fPhotoFlags('BAD_COUNTS_ERROR') != 0 AND
    dbo.fPhotoFlags('INTERP_CENTER') != 0 AND
    p.objid=s.objid and (specClass = 3 OR specClass = 4) AND 
    s.psfMag_i > 14.5 AND (s.psfMag_i - s.extinction_i) < 21.3 AND 
    s.psfMagErr_i < 0.2 
\end{verbatim}

\subsection{Experiment DR7b}
\begin{verbatim}
SELECT  
    s.specObjID, p.objid, p.ra, p.dec, s.targetObjID, s.z, s.zErr, 
    p.psfMag_u, p.psfMag_g, p.psfMag_r, p.psfMag_i, p.psfMag_z, 
    p.psfMagErr_u, p.psfMagErr_g, p.psfMagErr_r, p.psfMagErr_i, p.psfMagErr_z, 
    p.modelMag_u, p.modelMag_g, p.modelMag_r, p.modelMag_i, p.modelMag_z, 
    p.modelMagErr_u, p.modelMagErr_g, p.modelMagErr_r, p.modelMagErr_i, p.modelMagErr_z, 
    p.devMag_u, p.devMag_g, p.devMag_r, p.devMag_i, p.devMag_z, 
    p.devMagErr_u, p.devMagErr_g, p.devMagErr_r, p.devMagErr_i, p.devMagErr_z, 
    p.expMag_u, p.expMag_g, p.expMag_r, p.expMag_i, p.expMag_z, 
    p.expMagErr_u, p.expMagErr_g, p.expMagErr_r, p.expMagErr_i, p.expMagErr_z, 
    p.petroMag_u, p.petroMag_g, p.petroMag_r, p.petroMag_i, p.petroMag_z, 
    p.petroMagErr_u, p.petroMagErr_g, p.petroMagErr_r, p.petroMagErr_i, p.petroMagErr_z, 
    p.extinction_u, p.extinction_g, p.extinction_r, p.extinction_i, p.extinction_z, 
    p.devRad_u, p.devRad_g, p.devRad_r, p.devRad_i, p.devRad_z, 
    p.expRad_u, p.expRad_g, p.expRad_r, p.expRad_i, p.expRad_z, 
    p.petroRad_u, p.petroRad_g, p.petroRad_r, p.petroRad_i, p.petroRad_z, 
    p.petroR90_u, p.petroR90_g, p.petroR90_r, p.petroR90_i, p.petroR90_z, 
    p.petroR50_u, p.petroR50_g, p.petroR50_r, p.petroR50_i, p.petroR50_z, 
    p.devAB_u, p.devAB_g, p.devAB_r, p.devAB_i, p.devAB_z, 
    p.expAB_u, p.expAB_g, p.expAB_r, p.expAB_i, p.expAB_z 
    into mydb.qso_dr7_noflags from SpecPhoto as s, PhotoObjAll as p 

WHERE 
    p.SpecObjID = s.SpecObjID AND
    p.objid=s.objid and (specClass = 3 OR specClass = 4)
\end{verbatim}

\subsection{Experiment DR7+9}
\begin{verbatim}
SELECT 
   m.objid, m.ra AS ra1, m.dec AS dec1, 
   n.objid, n.distance, 
   p.ra AS ra2, p.dec AS dec2,
   p.objid, p.ra, p.dec, p.psfMag_u, p.psfMag_g, p.psfMag_r, p.psfMag_i, 
   p.psfMag_z,p.psfMagErr_u, p.psfMagErr_g, p.psfMagErr_r, p.psfMagErr_i, 
   p.psfMagErr_z,p.modelMag_u, p.modelMag_g, p.modelMag_r, p.modelMag_i, p.modelMag_z,
   p.modelMagErr_u, p.modelMagErr_g, p.modelMagErr_r, p.modelMagErr_i, 
   p.modelMagErr_z,p.devMag_u, p.devMag_g, p.devMag_r, p.devMag_i, p.devMag_z,
   p.devMagErr_u, p.devMagErr_g, p.devMagErr_r, p.devMagErr_i, p.devMagErr_z,
   p.expMag_u, p.expMag_g, p.expMag_r, p.expMag_i, p.expMag_z,p.expMagErr_u, p.expMagErr_g, 
   p.expMagErr_r, p.expMagErr_i, p.expMagErr_z,p.petroMag_u, p.petroMag_g, p.petroMag_r, 
   p.petroMag_i, p.petroMag_z,p.petroMagErr_u, p.petroMagErr_g, p.petroMagErr_r, 
   p.petroMagErr_i, p.petroMagErr_z,p.extinction_u, p.extinction_g, p.extinction_r, 
   p.extinction_i, p.extinction_z,p.devRad_u, p.devRad_g, p.devRad_r, p.devRad_i, 
   p.devRad_z,p.expRad_u, p.expRad_g, p.expRad_r, p.expRad_i, p.expRad_z,p.petroRad_u,
   p.petroRad_g, p.petroRad_r, p.petroRad_i, p.petroRad_z,p.petroR90_u, p.petroR90_g, 
   p.petroR90_r, p.petroR90_i, p.petroR90_z,p.petroR50_u, p.petroR50_g, p.petroR50_r, 
   p.petroR50_i, p.petroR50_z,p.devAB_u, p.devAB_g, p.devAB_r, p.devAB_i, p.devAB_z,p.expAB_u, 
   p.expAB_g, p.expAB_r, p.expAB_i, p.expAB_z
   into mydb.quasar_dr7_dr9_allphoto from MyDB.dr7_dr9_quasar AS m 
   
CROSS APPLY dbo.fGetNearestObjEq( m.ra, m.dec, 0.5) AS n 
JOIN PhotoObj AS p ON n.objid=p.objid
\end{verbatim}

\renewcommand{\arraystretch}{1.5}

\end{document}